\documentclass[aps, letterpaper, 11pt, onecolumn,
  floatfix, fleqn, notitlepage, 
  amsmath, amssymb, amsfonts,
  preprintnumbers, showpacs, showkeys,
  twoside]{revtex4-1}

\usepackage[colorlinks=true, linkcolor=blue, citecolor=blue,
            filecolor=blue, urlcolor=blue]{hyperref}
\usepackage{dcolumn, graphicx, moreverb}
\newcolumntype{.}{D{.}{.}{-1}} 
\usepackage[american]{babel}

\usepackage{siunitx}

\usepackage{mathptmx}
\usepackage[T1]{fontenc}

\usepackage{fancyhdr}
\fancypagestyle{empty}{%
  \fancyhf{}
  \fancyhead[LE,RO]{FERMILAB-TM-2598-AD-APC}
  \fancyfoot[C]{\thepage}
}

\newcommand{\bender}{\textit{bender}}
\newcommand{\func}[2]{#1\!\left(#2\right)}
\renewcommand{\d}{\mathrm{d}}
\newcommand{\beq}{\begin{equation}}
\newcommand{\eeq}{\end{equation}}

\begin{document}

\title{Field calculations, single-particle tracking, and beam dynamics with space charge in the electron lens for the Fermilab Integrable Optics Test Accelerator}
\thanks{Fermilab is operated by Fermi Research Alliance, LLC under contract no. DE-AC02-07CH11359 with the United States Department of Energy.}

\author{Daniel Noll}
\email{noll@iap.uni-frankfurt.de}
\affiliation{Institute of Applied Physics, Goethe University,
Max-von-Laue Str.~1, 60438 Frankfurt am Main, Germany}

\author{Giulio Stancari}
\affiliation{Fermi National Accelerator Laboratory, PO Box 500, Batavia, Illinois 60510, USA}

\date{\today}

\begin{abstract}
An electron lens is planned for the Fermilab Integrable Optics Test
Accelerator as a nonlinear element for integrable dynamics, as an
electron cooler, and as an electron trap to study space-charge
compensation in rings. We present the main design principles and
constraints for nonlinear integrable optics. A magnetic configuration of
the solenoids and of the toroidal section is laid out. Single-particle
tracking is used to optimize the electron path. Electron beam dynamics
at high intensity is calculated with a particle-in-cell code to estimate
current limits, profile distortions, and the effects on the circulating
beam. In the conclusions, we summarize the main findings and list
directions for further work.
\end{abstract}

\maketitle

\thispagestyle{empty}
\pagestyle{empty}

\newpage

\tableofcontents

\newpage

\section{Introduction}

Storage rings as well as cyclotrons traditionally rely on the linearity
of the particle optics, except where required to correct for effects
such as chromaticity. In these machines, the focusing depends linearly
on the offset of the particles from the axis and all particles oscillate
at approximately the same frequency, the tune. One limit on the
intensity of the beam in these machines is the maximum allowable spread
of tunes, which is typically limited by the presence of various
resonances.

To demonstrate the feasibility of a nonlinear optic in a real machine,
the Integrable Optics Test Accelerator (IOTA) ring is being built at
Fermilab. The idea is to produce a large tune spread using short
nonlinear insertions without reducing the region where beam can be
transported stably~\cite{nonlinearlattice}. The linac of the Fermilab
Accelerator Science and Technology (FAST) facility will be used as
injector~\cite{iotadesign}, generating a \SI{150}{\mega\eV} electron
beam.

\begin{figure}[b]
\begin{tabular}{cc}
\includegraphics[width=0.35\textwidth]{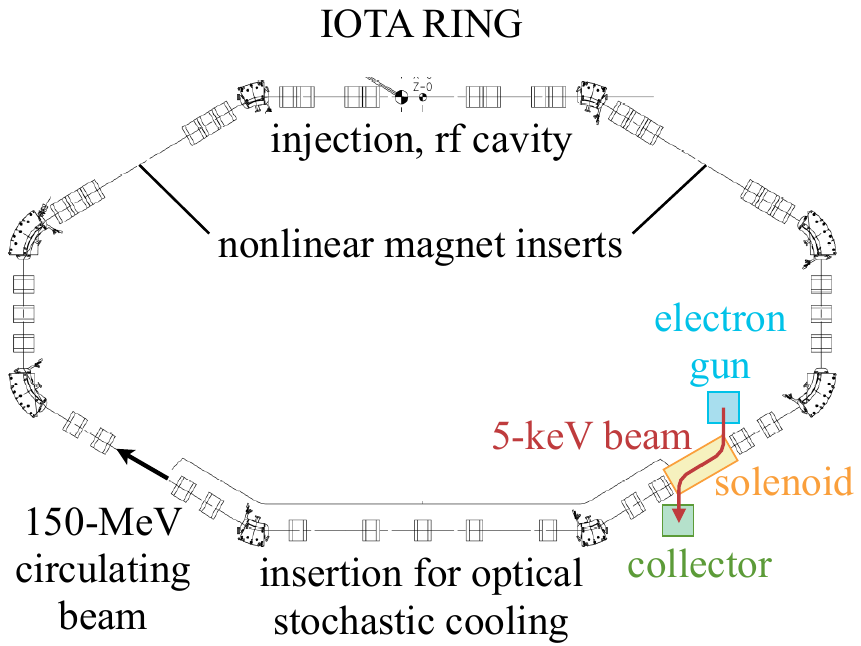} &
\includegraphics[width=0.6\textwidth]{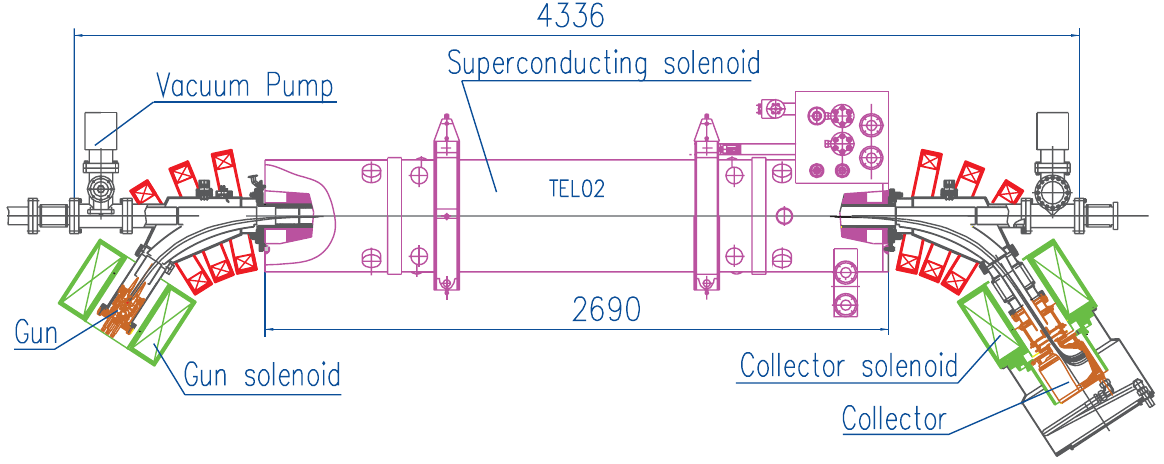} \\
Electron lens in IOTA &
Tevatron electron lens (TEL2)
\end{tabular}
\caption{Schematic layout of electron lenses.}
\label{fig:layouts}
\end{figure}

Electron lenses can be used to provide the fields required to make the
equations of motion integrable and thus non-chaotic. In an electron
lens, a low-energy electron beam is guided and compressed by magnetic
fields. The profile of this electron beam can be shaped to provide the
required kicks to the circulating beam~\cite{tevel2008}. Electron lenses
have been in operation at the Tevatron for compensation of beam-beam
effects and abort-gap cleaning~\cite{tevel2008} and are considered for a
number of other applications such as halo control in LHC
\cite{stancari_el}. Figure~\ref{fig:layouts} (left) shows a schematic
layout of the ring, which has a circumference of \SI{40}{\meter},
including the planned electron lens.

The schematic of one of the electron lenses in the Tevatron is shown in
Figure~\ref{fig:layouts} (right). The electron beam was produced by an
electron gun contained inside a solenoidal magnetic field on a potential
of \SIrange{-5}{-10}{\kilo\eV}. A short section consisting of three
coils provides a magnetic field to guide the electrons into the
superconducting main solenoid. An increase in magnetic field from the
gun solenoid (\SI{0.3}{\tesla} for TEL) to the main solenoid
(\SI{3.1}{\tesla}) compresses the electron beam by a factor of
\(\sqrt{B_{\mathrm{gun}}/B_{\mathrm{main}}}\). The main solenoid
included dipole correctors for alignment of the electron beam as well as
BPMs. The electron beam was then guided into a collector at a lowered
negative potential (to reduce power deposition) by a symmetric set of
coils.

For IOTA, two modi of operation are under investigation. In the first
case, the electric field of an electron beam of current \(I\) with the
radially symmetric current density
\beq
\func{j}{r} = \frac{I}{\pi}\frac{a^{2}}{\left(a^{2}+r^{2}\right)^{2}}
\eeq
generates two invariants of motion making all particle trajectories
regular and bounded. A one dimensional system like this was first
studied by McMillan and later extended to two dimensions
\cite{integrable_systems_round_coll_beams}. The electric field and
potential of such a distribution with an additional cut-off is listed in
Appendix~\ref{sec:mcmillan_dist}.

A second option is to use the main solenoid to set the beta function
\(\beta\) for the circulating beam to a constant value by using a
magnetic field of \(B_{z}=2\left(B\rho\right)/\beta\). In that case,
using any radially symmetric electron beam distribution will conserve
the Hamiltonian as well as the longitudinal component of the angular
momentum as long as the phase advance in the rest of the ring is a
multiple of \(\pi\)~\cite{stancari_el}.

Focus of this work is to find an initial design of the bending section,
which guides the electron beam into and out of the main solenoid, and to
investigate the dynamics of the high current electron beam using
particle-in-cell simulation tools~\cite{hockneyeastwood}. The beam
distribution i.e. its potential and electric field gained from these
simulations can then be used as input for long-time tracking simulation
of the beam in IOTA.

\section{Requirements for the bend design}

Instead of bending the beam in and out on the same side of the beam line
as in TEL2, the IOTA electron lens will feature bending section with
different signs of curvature. By doing this, the dipole kicks on the
circulating beam during the crossing of the bends should cancel. 

For the electron lens in IOTA, it was decided to reuse the electron gun
and collector solenoids from TEL2. Their coils have an inner diameter of
\SI{25}{\centi\meter} and an outer diameter of \SI{47.4}{\centi\meter},
as well as a length of \SI{30}{\centi\meter}~\cite{tel_magnetic_system}.
They can reach a field up to \SI{0.4}{\tesla} on axis. To reach the
required compression of the electron beam with a resistive main solenoid
however, only \SI{0.1}{\tesla} are required.

An electron energy of \SI{5}{\kilo\volt} is envisioned and is used
throughout this report.

\section{Geometry of the bending section}

\begin{figure}
    \centering
    \includegraphics[width=\textwidth,clip,trim=1.25cm 0 1.5cm 0]{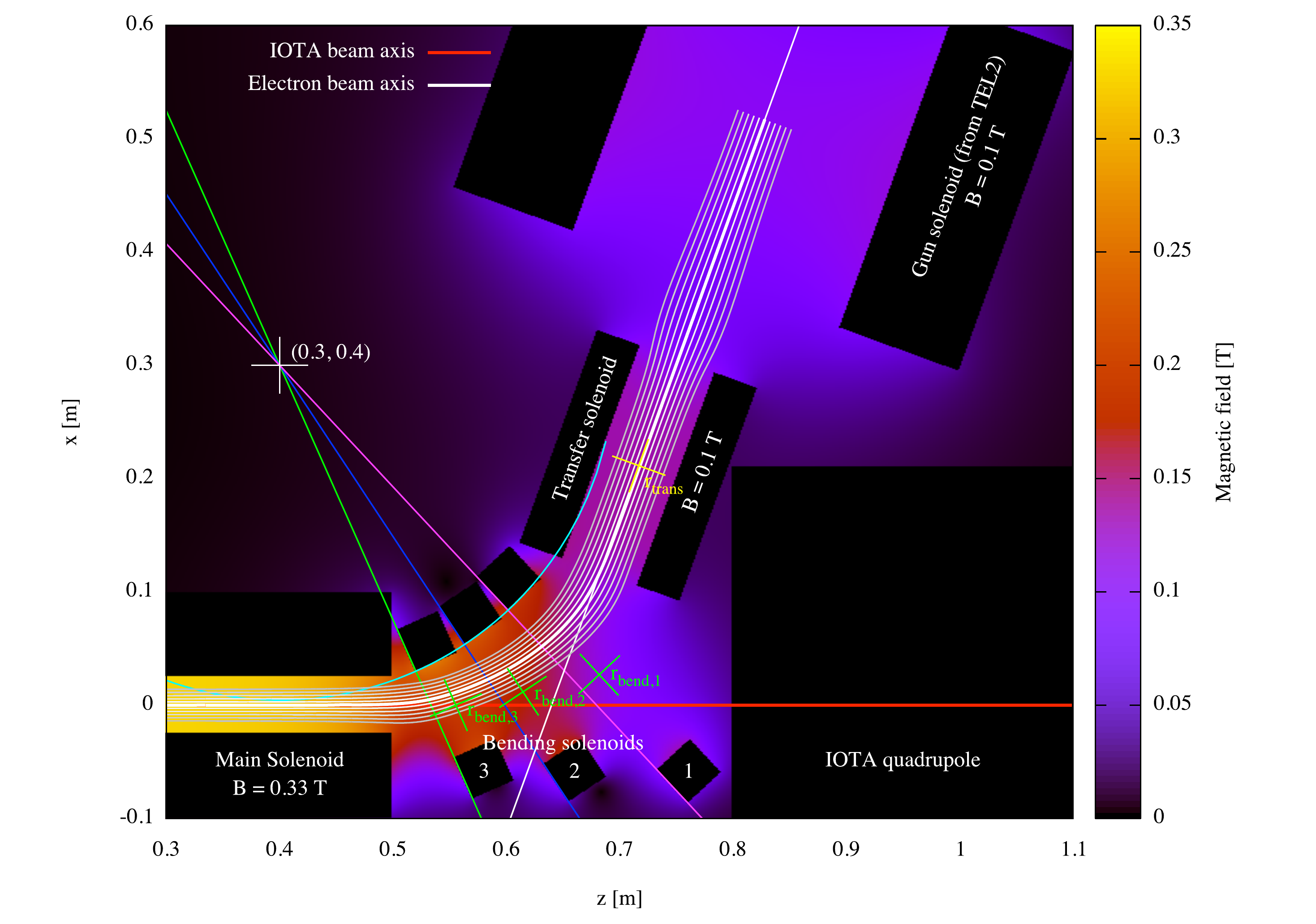}
    \caption{Schematic of an initial bend design for the IOTA electron lens.}
    \label{fig:iota_elens0_schematic}
\end{figure}

The distance of \SI{30}{\centi\meter} between the main solenoid and the
next IOTA quadrupole leaves little space for the bending solenoids.
Thus, the setup from the TEL2 electron lens had to be compressed as much
as possible. Figure~\ref{fig:iota_elens0_schematic} shows the setup that
was used for the following studies. All geometric parameters are
summarised in Table~\ref{tab:iota_elens0_free_parameters}. The
dimensions of the solenoids and coils used are presented in
Table~\ref{tab:iota_elens0_magnet_parameters}.

\begin{table}[t]
  \caption{Geometric parameters of the initial electron lens bend design.}
  \label{tab:iota_elens0_free_parameters}
  \centering
  \begin{tabular}{ l l l }
  \hline
    Center of circle for bending solenoid placement & \(z, x\) & \SI{40}{\centi\meter}, \SI{30}{\centi\meter}\\
    Radius of circle for bending solenoid placement & \(R_{\mathrm{bend,sol}} \) & \SI{29.56}{\centi\meter}\\
    Injection angle & \(\varphi_{\mathrm{inj}}\) & \SI{70}{\degree} \\
    z-axis crossing of straight line from injection & \(z_{\mathrm{inj}}\) & \SI{64}{\centi\meter}\\
    Distance between gun and transfer solenoid & \(d_{\mathrm{gun,transfer}}\) & \SI{7.5}{\centi\meter} \\
    \hline
  \end{tabular}   
\end{table}

\begin{table}[t]
  \caption{Parameters of the solenoids required for the initial electron lens bend design.}
  \label{tab:iota_elens0_magnet_parameters}
  \centering
  \begin{tabular}{ l | r | r | r | r | r | c }
  \hline
    Solenoid & \(R_{i}\) & \(R_{o}\) & \(B\) & \(j\) & \(\Theta\) & Center\\
             & [cm] & [cm] & [mT] & \(\left[\mathrm{A} / \mathrm{mm}^{2}\right]\) & \(\left[^\circ\right]\) & [m]\\\hline
    Gun      & 12.5 & 23.7 & 100 & 1.11 & 70 & (0.517, 0.828) \\
    Transfer & 3.5 & 7.5 & 100 & 2.27 & 70 & \(r_{\mathrm{trans}}=(0.2114, 0.717)\)\\
    Bend 1   & 9.65 & 13.65 & 53.9 & 6.28 & 43 & \(r_{\mathrm{bend,1}}=(0.02692, 0.682)\) \\
    Bend 2   & 6.33 & 10.33 & 74.9 & 6.28 & 33.5 & \(r_{\mathrm{bend,2}}=(0.0118, 0.615)\) \\
    Bend 3   & 4.17 & 8.17  & 100  & 6.28 & 24 & \(r_{\mathrm{bend,3}}=(0.0, 0.5555)\) \\
    Main     & 2.5  & 10    & 330  & 3.53 & & (0.0, 0.0) \\
    \hline
  \end{tabular}   
\end{table}

\begin{figure}
\begin{tabular}{cc}
\includegraphics[width=0.475\textwidth]{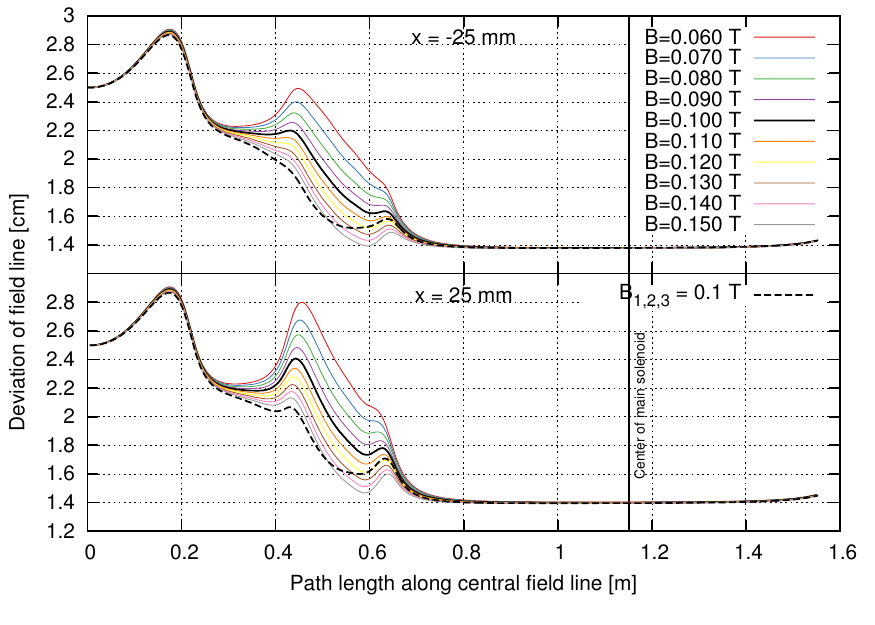} &
\includegraphics[width=0.475\textwidth]{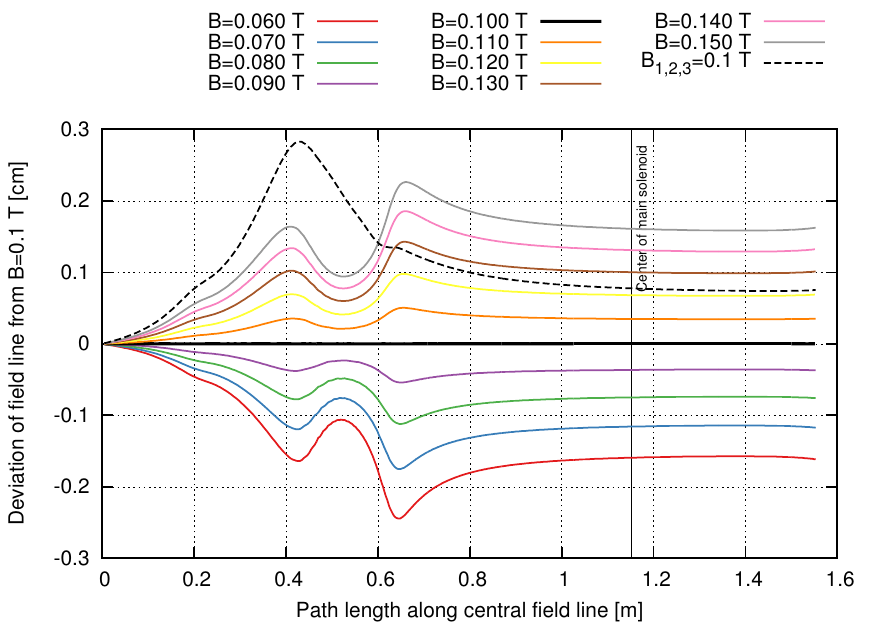}
\end{tabular}
\caption{Different settings of the magnetic field of the solenoid in
front of the main solenoid. Solid curves for the settings, where the
current through all solenoids are equal. Dashed curves for settings with
independent setting. Left: Offset of the field lines starting at the gun
at \SI{\pm25}{\milli\meter} from the central field line for different
settings of the magnetic field. Right: Offset of the central field line
compared to the one at \(B_{3}=\SI{0.1}{\tesla}\).}
\label{fig:iota0_sweep_bendsol_B}
\end{figure}

\begin{figure}
  \includegraphics[width=0.8\textwidth]
  {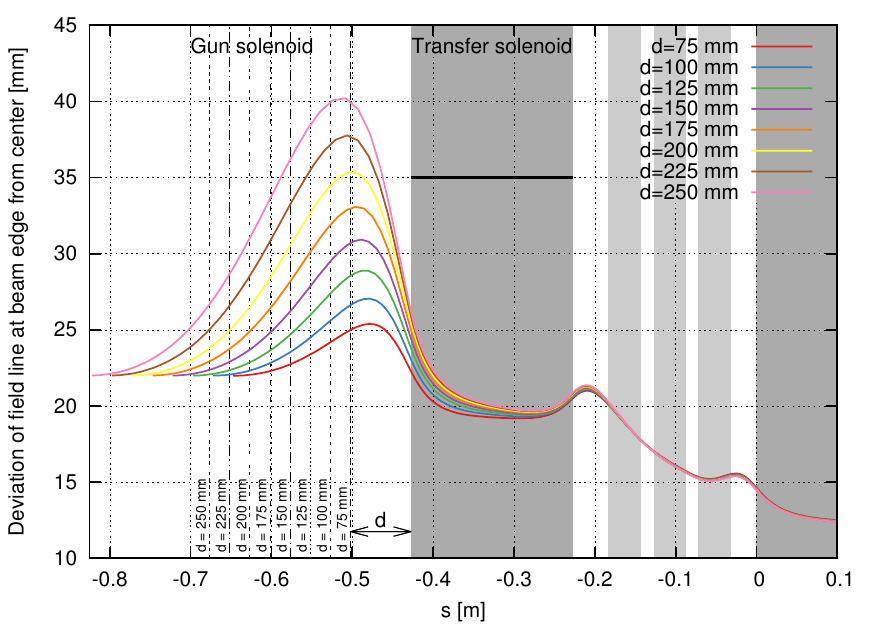}
  \caption{Field line divergence for different distances between gun and transfer solenoid.\label{fig:iota0_sweep_distance_transsol_gunsol}}
\end{figure}

\begin{figure}
\begin{tabular}{cc}
\includegraphics[width=0.475\textwidth]{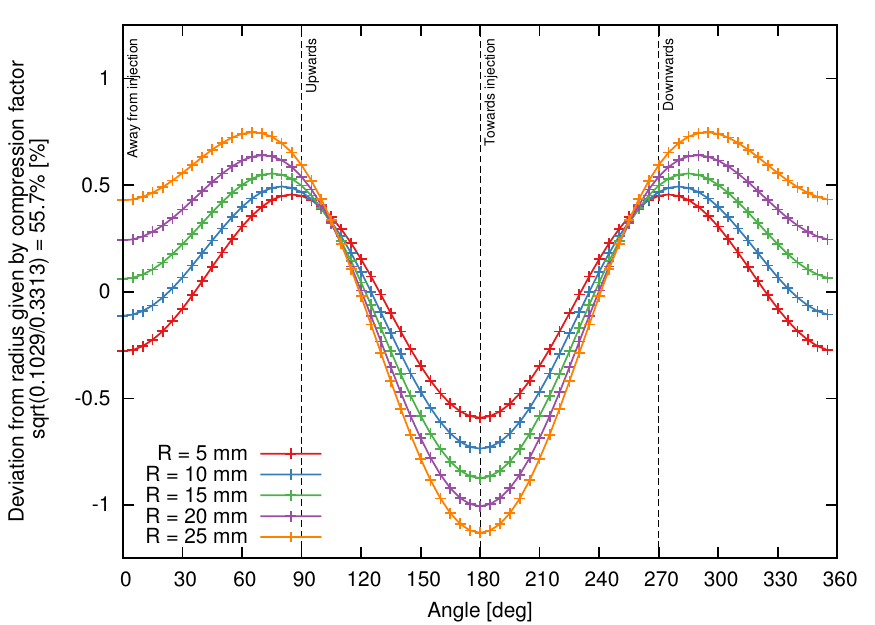} &
\includegraphics[width=0.475\textwidth]{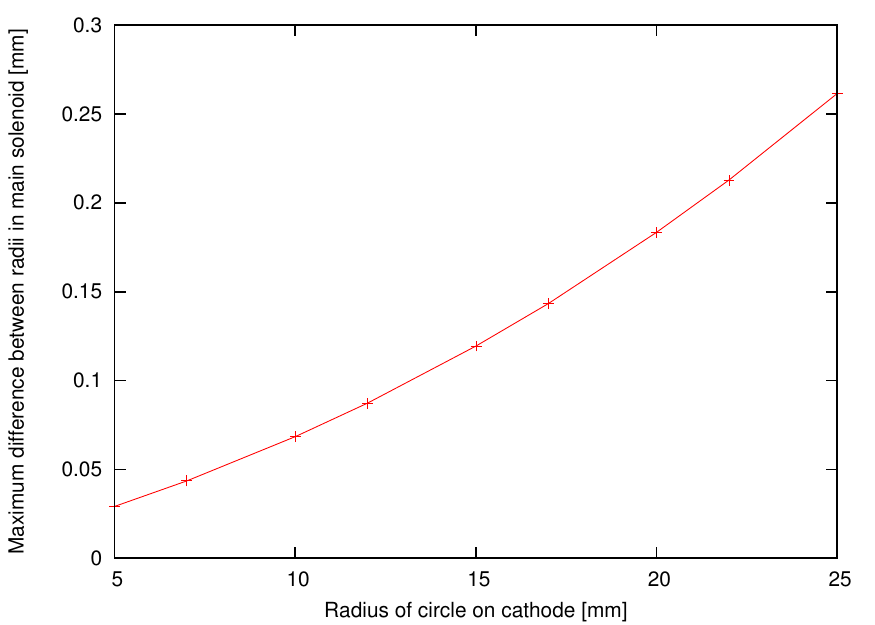}
\end{tabular}
\caption{Anisotropy of the magnetic field in the main solenoid. Left:
Angular dependency of deformation. Right: Maximal deformation of the
field lines.}
\label{fig:iota0_track_circle}
\end{figure}

The design was found using a simple Python code, tracking field lines by
numerical integration of
\beq
\frac{\d \vec{r}}{\d s} = \pm\frac{\vec{B}}{B}\mathrm{.}\label{eq:fieldline}
\eeq
For sufficiently large magnetic fields, it can be assumed, that the
electron beam will follow these lines with some gyration. The fields of
the solenoids were calculated by integrating the Biot-Savart law of
cylindrical conductors.

The inner, main solenoid-facing edges of the bending solenoids were
arranged on a circle of
\(R_{\mathrm{bend,sol}}=\SI{29.56}{\centi\meter}\). The inner diameters
of the bending solenoids were chosen, so that the outer, main-solenoid
facing edges touch the beam pipe. The origin of the circle as well as
the angles at which the three bending solenoids are placed were found by
trying to increase field homogeneity in the area of the beam.

To transfer the electron beam from the gun into the bending system an
additional solenoid is required. Together with the electron gun, this
solenoid was placed on an axis rotated by
\(\varphi_{\mathrm{inj}}=70^\circ\) which crosses the z-axis at
\(z_{\mathrm{inj}}=\SI{64}{\centi\meter}\).

Initially the currents through the bending solenoids were set to produce
the same field of \SI{0.1}{\tesla} on axis. This, however, led to high
current densities of over \SI{>10}{\ampere\per\square\milli\meter} for
the first and largest solenoid. To reduce the requirements and stay
below the value given as "good" in Ref.~\cite{irondominatedmagnets}, the
magnets were set to the same current. This reduces the field on axis for
the first solenoid to about \SI{54}{\percent} of the third and that of
the second one to about \SI{75}{\percent}, but provides the benefit of
operation with just one power supply. 

Additionally, reducing the field strength of these solenoids comes at
the cost of increased beam radius in the bends, although even at lower
field settings the beam should still be well separated from the vacuum
chamber. The difference for the outer field lines is shown in
Figure~\ref{fig:iota0_sweep_bendsol_B} (left). In all cases an asymmetry
between the inner and outer field lines can be seen in the bending
section, which can be as large as \SI{3}{\milli\meter}. This is a result
of the drop of magnetic field towards the center of bending solenoid 1.

Figure~\ref{fig:iota0_sweep_bendsol_B} (right) shows the deviation of the
central field line from the one at \(B=\SI{0.1}{\tesla}\) for different
coil currents. In addition to the difference in beam radii, there is an
offset from the axis of the solenoid as well.

In the design, a good way to influence the offset of the field lines in
the center of the main solenoid is to change the longitudinal position
of injection \(z_{\mathrm{inj}}\). Increasing \(z_{\mathrm{inj}}\) by a
centimeter increases the offset of the central field line by about
\SI{2.5}{\milli\meter}.

The distance \(d_{\mathrm{gun,transfer}}\) between the gun and transfer
solenoids can easily be increased to fit vacuum equipment. As can be
seen in Figure~\ref{fig:iota0_sweep_distance_transsol_gunsol}, the
fields lines bulge out more for increased values in the affected
section, but the influence on the bending section is already very small.
The maximum distance is thus limited by the available aperture between
the solenoids as well as the thickness of the beampipe inside the
transfer solenoid. However, there may be additional limits coming from
particle drifts and space charge, which will be investigated in the next
sections.

To quantify the anisotropy in the field, field lines starting on a
circle of radius \(R\) around the center of the cathode were tracked to
the center of the main solenoid. Assuming a perfectly homogeneous field,
the circle should shrink to
\(\sqrt{B_{gun}/B_{sol}}=\sqrt{0.1/0.33}=\SI{55.0}{\percent}\) of \(R\).
Stray fields increase the absolute field in the gun to
\SI{102.9}{\milli\tesla} and in the center of the main solenoid to
\SI{331.3}{\milli\tesla}. Figure~\ref{fig:iota0_track_circle} (left)
shows the relative deviation of circles of various radii from
\(\sqrt{102.9/331.3}R=\SI{55.7}{\percent} R\).

The deformation at low radii is mostly quadrupolar in nature. In
vertical direction, particles end up at a slightly increased distance
from the center, whereas particles in horizontal direction are guided
inwards more strongly. There is also a small difference between field
lines at positive and negative horizontal offset. The shift in the
maxima for higher \(R\) is a result of an outwards horizontal shift of
the field lines for both positive and negative vertical offsets.

The maximum deviation between field lines for the tracked circles is
shown in Figure~\ref{fig:iota0_track_circle} (right). For the field
lines on \(R=\SI{25}{\milli\meter}\) there is significant deformation of
up to \SI{0.26}{\milli\meter}.

\section{Single particle tracking}
\label{sec:sp_tracking}

\begin{figure}
    \centering
    \includegraphics[width=1.0\textwidth]{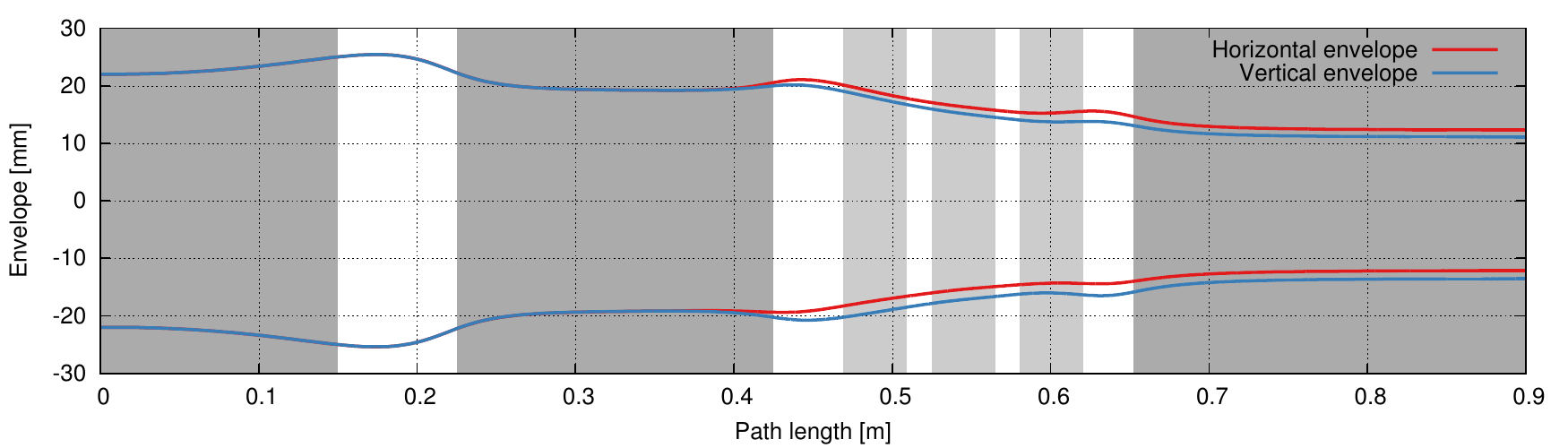}
    \caption{Envelope of a beam with radius \(r=\SI{22}{\milli\meter}\) and no emittance.}
    \label{fig:sp_beam_envelope}
\end{figure}

\begin{figure}
    \centering
    \includegraphics[width=1.0\textwidth]{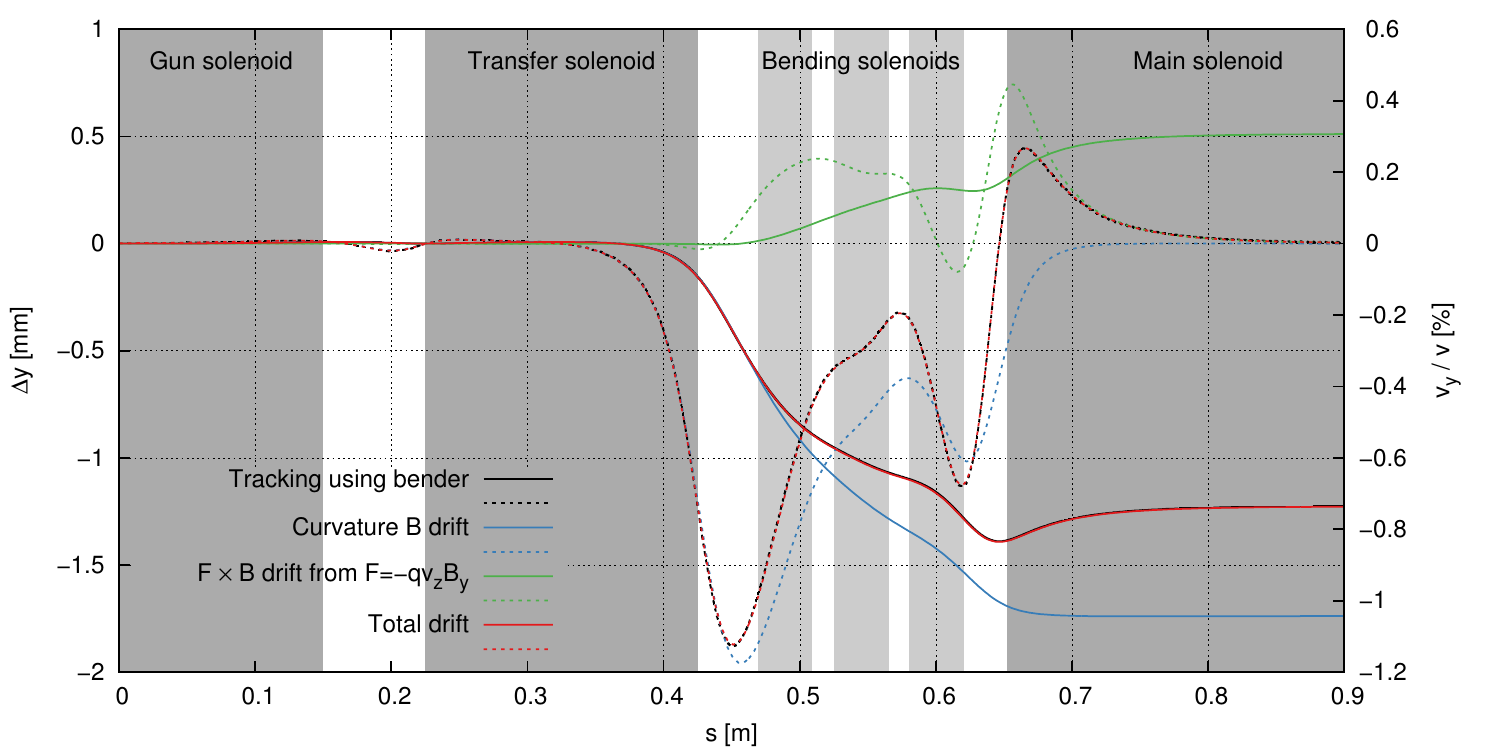}
    \caption{Particle drift along the central field line. Solid lines are vertical offsets, dashed lines are drift velocities. The drift is split in the influence of the curvature drift (blue) and the upward-guiding effect of the vertical magnetic field.}
    \label{fig:iota0_drift}
\end{figure}

\begin{figure}
  \centering
  \includegraphics[width=1.0\textwidth]{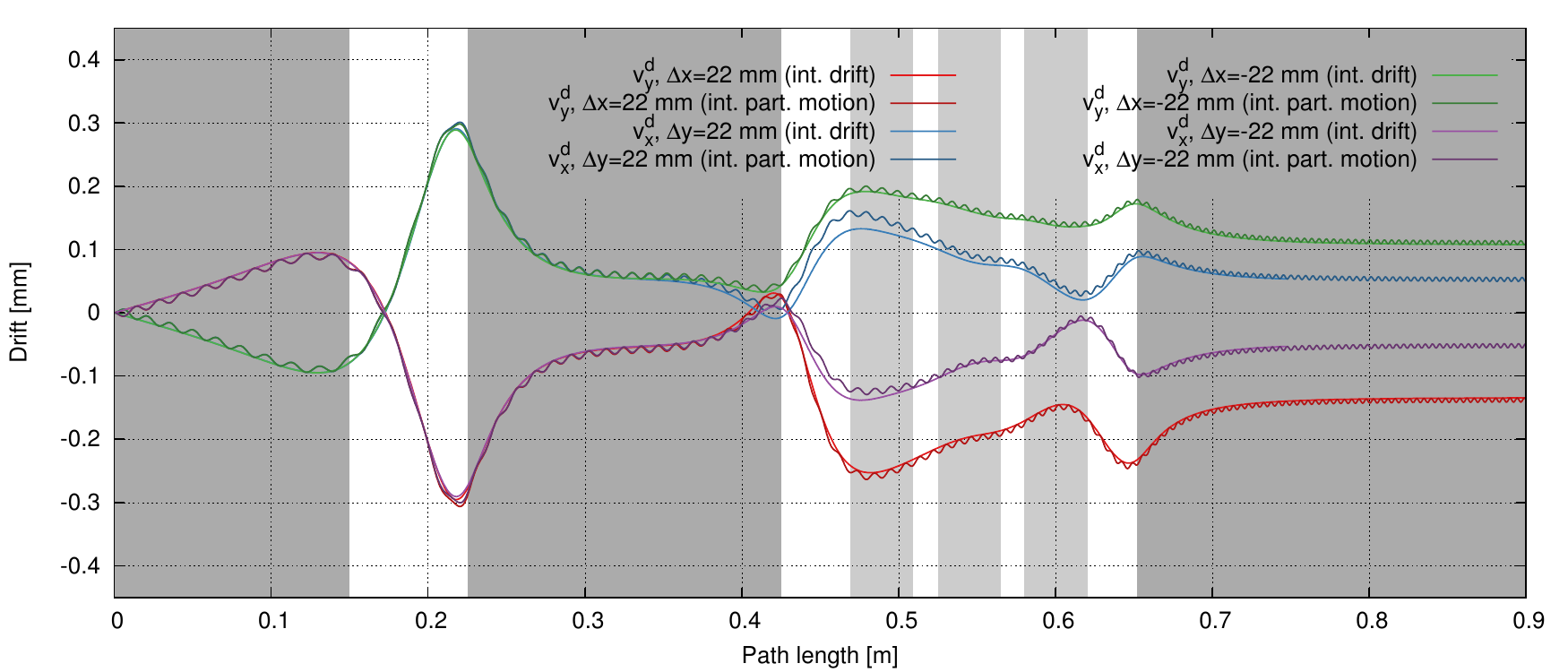}
  \caption{Horizontal and vertical drifts in reference to the central particle. Comparison between field line tracking using eq.  \ref{eq:fieldline_and_drift} ("integrated drift") and integration of motion using bender.\label{fig:transversal_drift_print}}
\end{figure}

\begin{figure}
  \centering
  \includegraphics[width=1.0\textwidth]{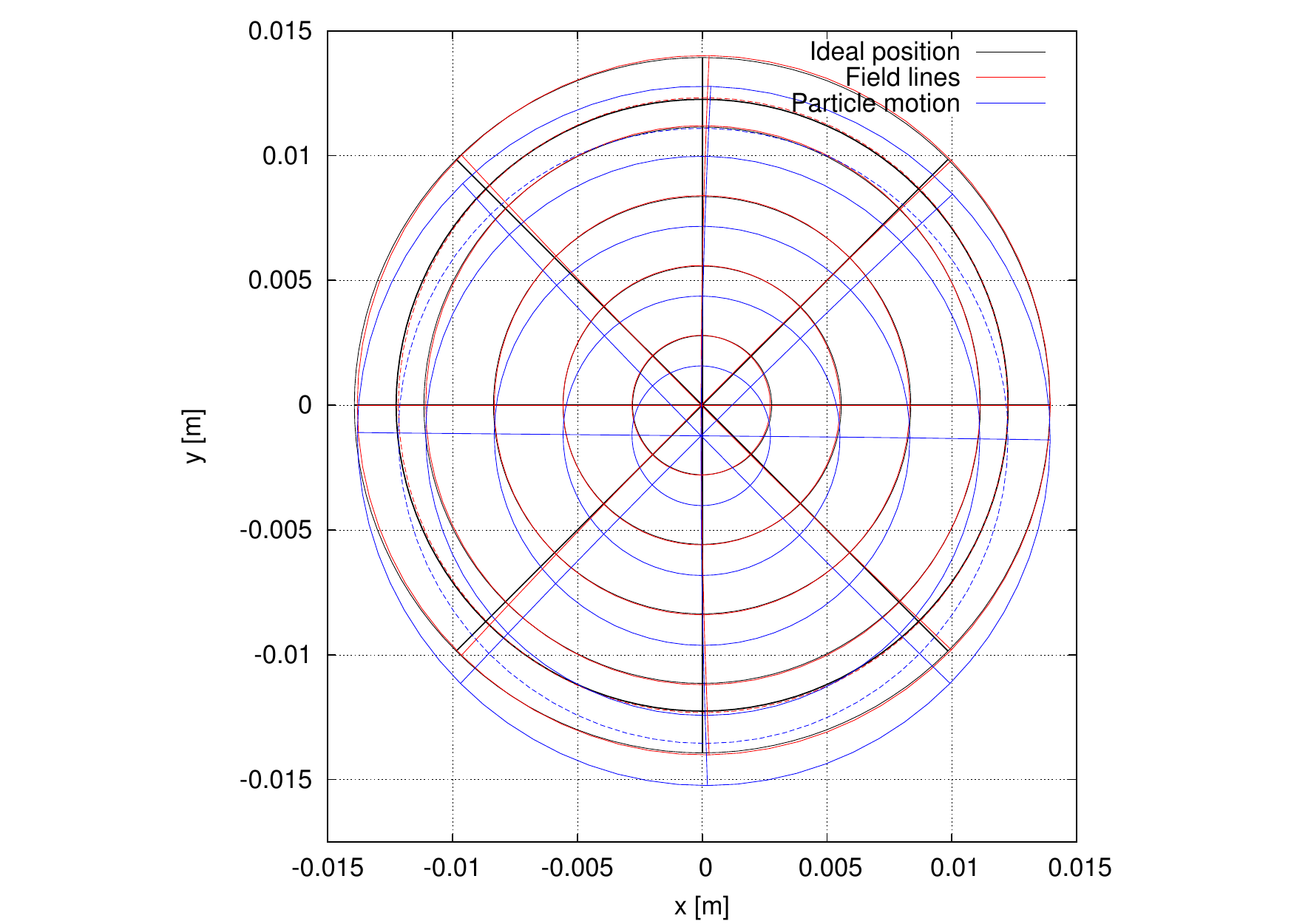}
  \caption{Graphical representation of the distortions of the trajectories in the center of the main solenoid from distortions in the field lines and from particle drifts. Coordinate system is in the reference frame of the electron beam. Dashed lines for particles starting with \(r=\SI{22}{\milli\meter}\) at the cathode.\label{fig:iota0_track_circle_deformation2d}}
\end{figure}

Single particles were tracked through the system using the
particle-in-cell code \bender{}~\cite{bender_ref}. Time steps of
\SI{5}{\pico\second} were used. A comparison with runs using smaller
steps of down to \SI{0.5}{\pico\second} showed no significant
differences.

To speed up particle tracking the solenoid fields were calculated on an
r-z grid with \SI{0.5}{\milli\meter} spacing and interpolated to the
particle positions. Comparisons to tracking simulations using the full
field showed only negligible differences.

Each solenoid contributes significantly to the magnetic field even at
large distances and inside the main solenoid. One influence of this is,
that the field line at the center of the gun solenoid is pointing at an
angle of \(\varphi'_{\mathrm{inj}}=\SI{70.097}{\degree}\), slightly
larger than \(\varphi_{\mathrm{inj}}\). For tracking, the source was
placed at this modified angle \(\varphi'_{\mathrm{inj}}\), to minimize
the initial magnetic moment of the particles as well as the cyclotron
motion. \(\varphi'_{\mathrm{inj}}\) also varies significantly over the
cathode region in horizontal direction, for example increasing linearly
from \SI{69.98}{\degree} to \SI{70.22}{\degree} in the range of
\(x=\SI{\pm22}{\milli\meter}\).

Furthermore, it was observed that truncating the fields of any of the
solenoids inside the simulation volume will have an influence on
particle trajectories. Thus, a grid for each magnet spanning the whole
system was used.

Figure~\ref{fig:sp_beam_envelope} shows the beam envelope of a
\SI{5}{\kilo\electronvolt} electron beam with negligible emittance. The
compression by \SI{55.7}{\percent}, i.e. from \SI{22}{\milli\meter} to
\SI{12.25}{\milli\meter}, is equal to the compression of the field
lines. The assumption, that the electrons follow the field lines, seems
to be fulfilled well. The shift downwards is a result of particle
drifts, which will be investigated in the following.

In a magnetic guiding field, when a force acts on a charged particle, it
will drift in a direction perpendicular to this force and the magnetic
field. The centrifugal force in a bend leads to a curvature drift
\beq
  \vec{v}_{R} = \frac{m v_{\parallel}^{2}}{q R_{c}}\frac{\vec{R}_{c}\times\vec{B}}{B^{2}}\,\mathrm{,}\label{eq:curvbvelocity}
\eeq
where \(\vec{R}_{c} = \d^{2}\vec{r}/\d s^{2}\) is pointing in the
direction of the normal vector along the trajectory and has the
magnitude of the radius of a circle that fits the trajectory at each
given point. Assuming \(\vec{R}\) and \(\vec{B}\) to be perpendicular,
\(B\) pointing only in longitudinal direction and \(R\) only in
horizontal direction along the particle trajectories, results in a
vertical drift. Integrated over path length, the total drift is
\beq
\func{\Delta y}{s} = \frac{2U}{v_{0}}\frac{q}{\left|q\right|}\int_{0}^{s}\d s' \frac{1}{\func{R}{s'}\func{B}{s'}}\,\mathrm{.}
\eeq
The curvature B drift was responsible for several millimeters of
vertical drift in TEL-1~\cite{tevel2008}.

For the design of the bends presented above, the curvature B drift is as
large as \SI{1.7}{\milli\meter}. As the particle moves downwards, it
moves into a region where the magnetic field, especially from the main
solenoid, is pointing upwards. This vertical field will then guide the
particle upwards again. Assuming that particles follow the field lines
exactly:
\beq
 \frac{\d y}{\d s} = \frac{B_{y}}{B_{z}} \quad\mathrm{and}\quad \func{\Delta y}{s} = \int_{0}^{s} \d s' \frac{\func{B_{y}}{s'}}{\func{B_{z}}{s'}}\label{eq:fieldlineguiding}
\eeq

Interestingly, the same equation can also be derived by thinking of the
upward movement as an \(\vec{F}\times\vec{B}\) drift. The vertical
magnetic field produces a force mainly in horizontal direction \(F_{x} =
-q v_{z} B_{y}\). This force while being negligible in horizontal
direction produces a vertical drift, with the velocity
\beq
v_{y} = \frac{1}{q}\frac{\vec{F}\times\vec{B}}{B^{2}} =
-\frac{v_{z}B_{y}B_{z}}{B^{2}}\vec{e}_{x}\times\vec{e}_{z} =
v_{z}\frac{B_{y}B_{z}}{B^{2}} \overset{B_{z}\approx B}{=}
v_{z}\frac{B_{y}}{B_{z}}\,\mathrm{,}\label{eq:curvbdrift}
\eeq
which is identical to eq. \ref{eq:fieldlineguiding}.

To follow the drifts along with the field line,
\beq
\frac{\d \vec{r}}{\d s} = \pm\frac{1}{B}\left\{\vec{B} + \frac{m}{q
v_{0}}\frac{1}{B^{2}}\left(v_{||}^{2} +
\frac{1}{2}v_{\perp}^{2}\right)\left(\vec{B}\times\vec{\nabla}
B\right)\right\} \label{eq:fieldline_and_drift}
\eeq
instead of eq. \ref{eq:fieldline} can be used. The grad B drift can be
neglected by setting \(v_{\perp}=0\).

The deviation on the central field line resulting from both drifts is
plotted in Figure~\ref{fig:iota0_drift}. The total offset for the central
particle in vertical direction is \SI{-1.2}{\milli\meter}. The
trajectories curvature has two maxima of similar dimensions at the start
and the end of the first and the third bending solenoid. Due to the
higher magnetic field closer to the main solenoid, the region between
transfer and the first bending solenoid accounts for most of the
downwards curvature B drift.

For particles with an offset from the central field line, there is an
additional drift that is a result from the compression of the beam. On
these trajectories \(\vec{R}\) is not confined to the bending plane any
more, but -- in the frame of reference relative to the central field
line -- also acquires a component pointing radially. Because of
\(\vec{e}_{r}\times\vec{e}_{z}=-\vec{e}_{\varphi}\), this component
causes particles to drift from their position in an angular direction.

Figure~\ref{fig:transversal_drift_print} shows the drift of particles
with an offset of \SI{\pm22}{\milli\meter} in \(x\) and \(y\) in the
perpendicular transverse direction in the frame of reference of the
central particle\footnote{For correctness: for the particles with
vertical offset, the reference frame follows the field line at the
particle's initial position to correct for the horizontal distortion of
field lines starting at a vertical offset mentioned above.}. In the
injection part of the system, the field and thus the drift is
approximately rotationally symmetrical around the beam axis.

Because there are no differences in absolute field between particles,
the drift velocities follow the curvature of the trajectories.
Initially, the curvature has a small negative value (radial vector
\(\vec{R}\) pointing inwards), leading to an initial drift upwards for
particles with positive \(\Delta x\). About \SI{5}{\centi\meter} before
the end of the gun solenoid curvature changes sign and increases to
about 10 times of the initial value, which results in a much higher
drift in the opposing direction. This drift is then reduced because of
another change in the sign of the curvature at the start of the transfer
solenoid. Most of the drift is canceled in the center of the transport
solenoid, only about \(\Delta r=\SI{0.05}{\milli\meter}\) remains.

In the bending section, due to an horizontal gradient in the magnetic
field, the magnitude of the vertical drift differs from that in
horizontal direction. The qualitative behavior however is still
symmetric between the x and y planes, i.e. the drift still follows the
curvature produced by the compression of the field lines.

The difference in magnetic field for particles starting with
\SI{\pm22}{\milli\meter} horizontal offset can be as high as
\SI{50}{\milli\tesla}. This means that the drift produced by the
curvature from compression of the field lines will have a smaller
influence on particles with negative offset (higher magnetic field) than
on those with positive offset. This is visible in
Figure~\ref{fig:transversal_drift_print} in the smaller variation of the
(green) curve for the particle with \(\Delta x=\SI{-22}{\milli\meter}\)
in comparison to the particle starting with \(\Delta
x=\SI{22}{\milli\meter}\) (red curve).

In addition, the horizontal gradient in the magnetic field also changes
the total downward drift of the particles. The higher the magnetic
field, the lower the drift resulting from the curvature produced by the
bending. For this reason, particles on the inner trajectories at higher
magnetic field experience a lower total vertical downward drift and thus
are moving upwards in reference to the central particle, whereas
particles on the outer trajectories are moving downwards. This effect
adds to the drift resulting from the compression, producing a larger
drift in Figure~\ref{fig:transversal_drift_print} in vertical direction
than in horizontal direction.

Figure~\ref{fig:iota0_track_circle_deformation2d} summarizes all the
distortions of the particle trajectories.

\section{Beam dynamics with space charge}

Apart from providing kicks to the propagating beam in IOTA, the electric
field of the electron lens beam also influences the beam itself. To
understand the dynamics and estimate the influence of space charge,
simulations using \bender{} were made.

\bender{} simulates DC beams by injecting slices of macroparticles into
the simulation in every time step. For all simulations presented below,
\(\Delta t=\SI{5}{\pico\second}\) was used. For the electron lens, using
100 particles inserted per step results in a total number of 430k
particles. During beam formation, particles at the head of the beam are
accelerated away from the particles being injected behind them -- to up
to 4 times their initial energy. Depending on the beam current, this
switch-on effect leads to visible deviation for up to twice the transit
time of the beam. For all plots presented here, these particles were
discarded.

As an electrostatic code, \bender{} only considers the self electric
fields of the beam by computing them numerically from Poissons equation,
but neglects the self magnetic fields. This approximation should pose no
problem in this case, since the magnetic field produced by the electron
beam, for example \(B=\mu_{0}I / (2\pi
R_{\mathrm{beam}})=\SI{20}{\micro\tesla}\) for \(I=\SI{2}{\ampere}\) and
\(R_{\mathrm{beam}}=\SI{2}{\centi\meter}\), is in the same order of
magnitude as earth's magnetic field and thus much lower than the guiding
fields.

The electron gun was not included in the simulation. Instead, the beam
was assumed to have the required distribution. The potential on the
bounding surface was fixed to the analytic expression of a 2d infinite
beam. Since the beam size in the gun solenoid does not vary much, this
approximation should be valid. In future simulations however, the
electron gun or a distribution from an additional simulation of the gun
should be included.

Boundary conditions were set on cylinders at the inner diameters of the
coils. For the bend, a tube with a radius of \SI{3}{\centi\meter} was
put on a circle connecting the end of the transfer solenoid and the
start of the main solenoid using CAD software and included using the STL
import in \bender{}. The influence of the shape of this pipe should be
studied in more detail to provide input for the technical design.

"Low" resolution runs were made using a grid resolution of
\SI{1}{\milli\meter} grid spacing in all directions (\num{3.6e6} grid
points total) on the Wilson cluster~\cite{wilson_cluster}. On 16
processors of the amd32 partition, a simulation of 15000 steps took
about \SI{20}{\hour}. The processors were distributed by splitting the
domain in an upper and lower half and by putting 4 domains on the IOTA
beam pipe and 4 following the bends for each half.

"High" resolution runs were made on 192 processors and a grid spacing of
\SI{0.5}{\milli\meter} (\num{21.3} million grid points) at 2500
particles inserted per step (10 million particles in total), taking
\SI{9}{\hour} for 15000 steps on the intel12 partition of Wilson. For
these, the parallel particle-grid accumulation routine in \bender{} had
to be reimplemented.

\subsection{Particle dynamics and current limits}

Assuming radial symmetry of the beam, the electric field will only point
in radial direction in the absence of compression. An electron which
starts with negligible transverse velocity at the cathode will thus be
accelerated outwards. However, as the electron gathers transverse
momentum, the magnetic force will increase and transfer momentum from
the radial direction into the angular direction. After some outward
acceleration, the outwards pointing electric and the inwards-pointing
magnetic force will equal the centripetal force on the electron. At this
point, it will start to move inwards again until it ends up at the
radius it started out at, but with a slightly increased angle in respect
to the axis.

This motion is the source of the \(E\times B\) drift in the angular
direction. Looking at the case of a homogeneous beam, its frequency and
revolution time are
\beq\omega_{E\times B}=\frac{1}{2\pi\epsilon_{0}}\frac{I}{v}\frac{1}{R_{\mathrm{b}}^{2}}\frac{1}{B}\quad\mathrm{and}\quad T=\frac{4\pi^{2}\epsilon_{0} v B R_{\mathrm{b}}^{2}}{I}\mathrm{.}\eeq
For a homogeneous beam, the rotation frequency does not depend on \(r\)
and thus is equal for all particles. Ignoring longitudinal effects,
since \(\func{R_{\mathrm{b}}}{s}=\sqrt{B_{0}/\func{B}{s}}R_{0}\), the
oscillation frequency is also independent of s, because
\beq\omega_{E\times B}=\frac{1}{2\pi\epsilon_{0}}\frac{I}{v}\frac{1}{R_{\mathrm{0}}^{2}}\frac{\func{B}{s}}{B_{0}}\frac{1}{\func{B}{s}}=\frac{1}{2\pi\epsilon_{0}}\frac{I}{v}\frac{1}{R_{\mathrm{0}}^{2}}\frac{1}{B_{0}}\mathrm{.}\eeq
A \SI{2}{\ampere} beam of \SI{2}{\centi\meter} radius will rotate with a
frequency of only \SI{3.4}{\mega\hertz} around its own axis, by
approximately \SI{33}{\degree} from the cathode to the center of the
main solenoid. The value as well as the behaviour is matched by the
\bender{} simulation by about \SI{1}{\degree}\footnote{Without
subtracting the additional particle drifts.}.

The distance an electron at radius \(r_{0}\) will travel outwards before
it is reflected is given by
\beq
r=\frac{1}{\epsilon_{0}}\frac{m}{e}\frac{I}{v}\frac{1}{\pi R_{\mathrm{b}}^{2}}\frac{1}{B^{2}}r_{0}\mathrm{.}\label{eq:spsc_excursion}
\eeq
For the parameters given above, the deviation of particles at the edge
of the beam at the cathode (B=\SI{0.1}{\tesla}) is only
\SI{49}{\micro\meter}. Thus, for homogeneous distributions, no
significant deformation will occur. 

The current limit for a similar system without any bending or beam
compression is given by the Brillouin flow limit~\cite{davidson_nnp},
\beq I_{\mathrm{brillouin}} < \frac{1}{2}\frac{e\epsilon_{0}}{m}v_{\mathrm{b}} \pi R_{\mathrm{b}}^{2} B^{2} \approx \SI{410}{\ampere}\mathrm{.} \eeq
For currents larger than \(I_{\mathrm{brillouin}}\), the magnetic force
is not able to balance the electric force any more. For a real electron
lens, the current is thus only limited by longitudinal effects. Ignoring
possible distortions in the beam shape, the individual slowing of
particles at different radii as well as the beams emittance, this
current limit is given by
\beq
I < 4\pi\epsilon_{0}\sqrt{\frac{e}{m}}U^{3/2}\left(\frac{g_{1}-\sqrt{3 g_{2}^{2} - g_{1}^{2}/3}}{4g_{1}^{2}-9g_{2}^{2}}\right)\mathrm{,}
\eeq
with the geometric factors \(g_{1}=\func{\ln}{R_{\mathrm{b}} /
R_{\mathrm{p1}}}-0.5\) and \(g_{2}=\func{\ln}{c R_{\mathrm{b}} /
R_{\mathrm{p2}}}-0.5\). For derivation of this formula, see Appendix~\ref{sec:app_currentlimit}. For
\(E_{\mathrm{b}}=\SI{5}{\kilo\electronvolt}\), \(c=0.55\),
\(R_{\mathrm{p1}}=\SI{4}{\centi\meter}\) and
\(R_{\mathrm{p2}}=\SI{2.5}{\centi\meter}\) the current limit is
\SI{7.9}{\ampere} (\SI{9.2}{\ampere} from the numerical solution). From
the simulation, the current limit for a homogeneous beam of
\(R_{\mathrm{b}}=\SI{2}{\centi\meter}\) seems to be between
\SI{7.5}{\ampere} and \SI{10}{\ampere}. At these currents, a significant
build-up of charge somewhere around the transfer solenoid can be
observed, which produces a potential large enough to deflect some of the
incoming electrons.

For the McMillan case, the rotation frequency depends on the radius,
\beq\omega_{E\times B}=\frac{1}{2\pi\epsilon_{0}}\frac{I}{v}\frac{1}{B}\left(1+\frac{a^{2}}{c^{2}}\right)\frac{1}{r^{2}+a^{2}}\mathrm{.}\eeq
For the \SI{1.7}{\ampere} case, oscillation frequencies (revolution
times) range from \SI{2.4}{\mega\hertz} (\SI{417}{\nano\second}) on the
edge of the beam to \SI{92}{\mega\hertz} (\SI{10.9}{\nano\second}) near
the beam center.

The field in the core of a McMillan distributed beam is also linear,
although increased by a factor of \((R_{\mathrm{b}}/a)^{2}+1\) compared
to a homogeneously distributed beam with radius \(R_{\mathrm{b}}\). For
a particle at \(r=\SI{1}{\milli\meter}\) in a \SI{1.7}{\ampere} McMillan
distributed beam for example, the excursion predicted by eq.
\ref{eq:spsc_excursion} is \SI{200}{\micro\meter} or \SI{20}{\percent}.
This means, that for the McMillan distribution at high currents some
deviation in the profiles can be expected.

\subsection{Profiles}

\begin{figure}
\begin{tabular}{cc}
\includegraphics[height=0.475\textwidth, angle=-90]{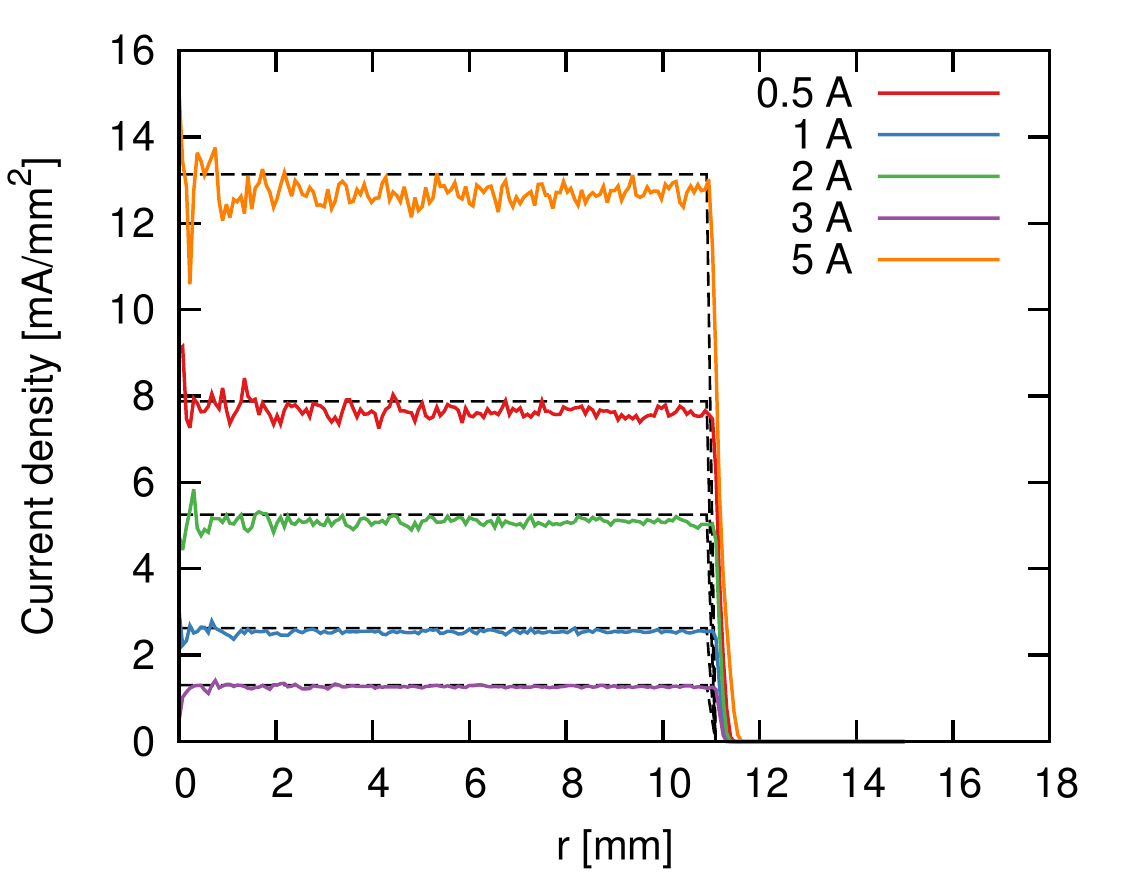} &
\includegraphics[height=0.475\textwidth, angle=-90]{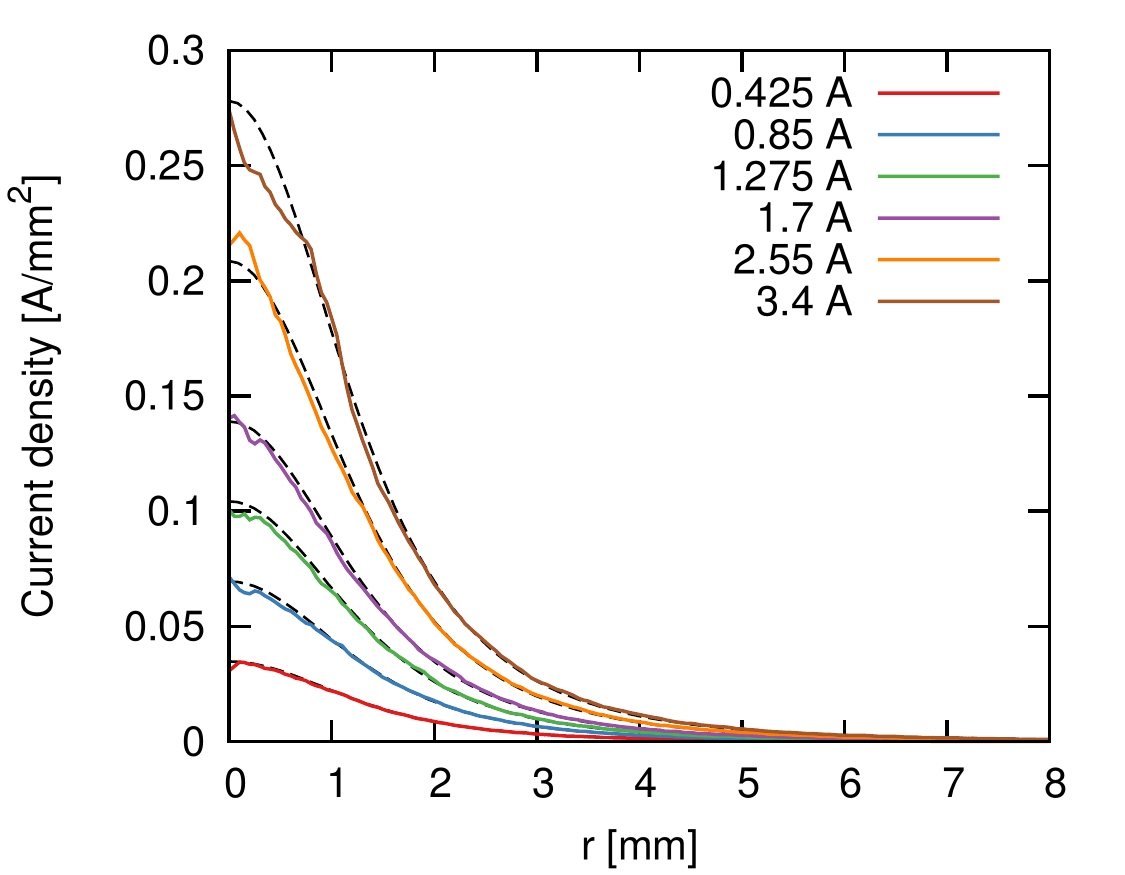}
\end{tabular}
\caption{Current density profiles at \(z=\SI{30}{cm}\) for various beam
currents from the "low resolution" simulation. Left: Homogeneous beam.
Right: McMillan distributed beam. The offset from the drift was
subtracted before calculation of the current density.}
\label{fig:iota_sc_profile}
\end{figure}

Figure~\ref{fig:iota_sc_profile} shows the current density profiles in
the electron lens. For the homogeneous distribution (left), there is a
small systematic dependence of the beam size on the current.

For the McMillan distribution (Figure~\ref{fig:iota_sc_profile}, right),
there is a clear deviation from the required profile for the
\SI{3.4}{\ampere} simulation: the current density in the core is
slightly reduced. The profiles of simulations with lower current follow
the theoretical distribution in the tail, but tend to be systematically
lower than the theoretical current density around
\(r=\SI{1}{\milli\meter}\). In the core of the distribution, there is no
systematic behaviour when current increases, probably due to an
insufficiently high number of particles close to the axis.

It should be noted, that the grid resolution of the simulation was far
below the the deviations seen in these results, making it difficult to
draw a final conclusion.

Table~\ref{tab:current_distribution_deviation} contains the deviation of
the profiles from the simulations of the transport of the McMillan
distributed beam. The rms deviation increase from about 3\% for the
simulation without space charge to 13\% for the visibly distorted
profile at \SI{3.4}{\ampere}. For the two current values where  high
resolution simulations where made, the rms values differ by 1.4\%
(\SI{2}{\ampere}) and 0.3\% (\SI{3.4}{\ampere}) from the lower
resolution one respectively. The maximum error values show a significant
influence on the resolution of the transport simulation and also don't
increase systematically. It was observed, that changing the resolution
on which these errors are calculated significantly changes the value of
the residuals.

\begin{table}[t]
  \caption{Maximum, arithmetic and rms deviation of the profiles extracted from the simulation to the expected McMillan distribution. The errors were calculated using the same grid resolution used for the high resolution simulation. \(N\) is the number of points the current density was calculated on and \(j_{0}\) the peak current density.}
  \label{tab:current_distribution_deviation}
  \centering
  \begin{tabular}{| l | r | r | r | r | r | r |} \hline
 & \(\max(\delta)\) & \(N^{-1}\sum\delta\) & \(\sqrt{N^{-1}\sum\delta^{2}}\) & \(j_{0}^{-1}\max(\delta)\) & \(j_{0}^{-1}N^{-1}\sum\delta\) & \(j_{0}^{-1}\sqrt{N^{-1}\sum\delta^{2}}\)\\
 & \multicolumn{3}{c |}{[\si{\kilo\ampere\per\meter\squared}]} & \multicolumn{3}{c |}{ [\%] }\\\hline
no space charge & & & & 3.9 & < 0.1 & 2.9\\
\SI{0.425}{\ampere} & 2.9 & 0.029 & 1.9 & 8.3 & 0.1 & 5.5 \\
\SI{0.85}{\ampere} & 7.6 & 0.069 & 4.5 & 10.9 & 0.1 & 6.4 \\
\SI{1.275}{\ampere} & 14.1 & 0.11 & 7.3 & 13.4 & 0.1 & 7.0 \\
\SI{1.7}{\ampere} & 16.7 & 0.17 & 10.8 & 12.0 & 0.1 & 7.8 \\
\SI{1.7}{\ampere} (high resolution) & 12.1 & 0.14 & 9.0 & 8.7 & 0.1 & 6.4\\
\SI{2.55}{\ampere} & 26.6 & 0.31 & 20.1 & 12.7 & 0.1 & 9.6 \\
\SI{3.4}{\ampere} & 60.2 & 0.54 & 35.2 & 21.6 & 0.2 & 12.6\\
\SI{3.4}{\ampere} (high resolution) & 41.0 & 0.55 & 36.1 & 14.7 & 0.2 & 12.9 \\
\hline
  \end{tabular}   
\end{table}

\begin{figure}
  \centering
  \includegraphics[width=1.0\textwidth]{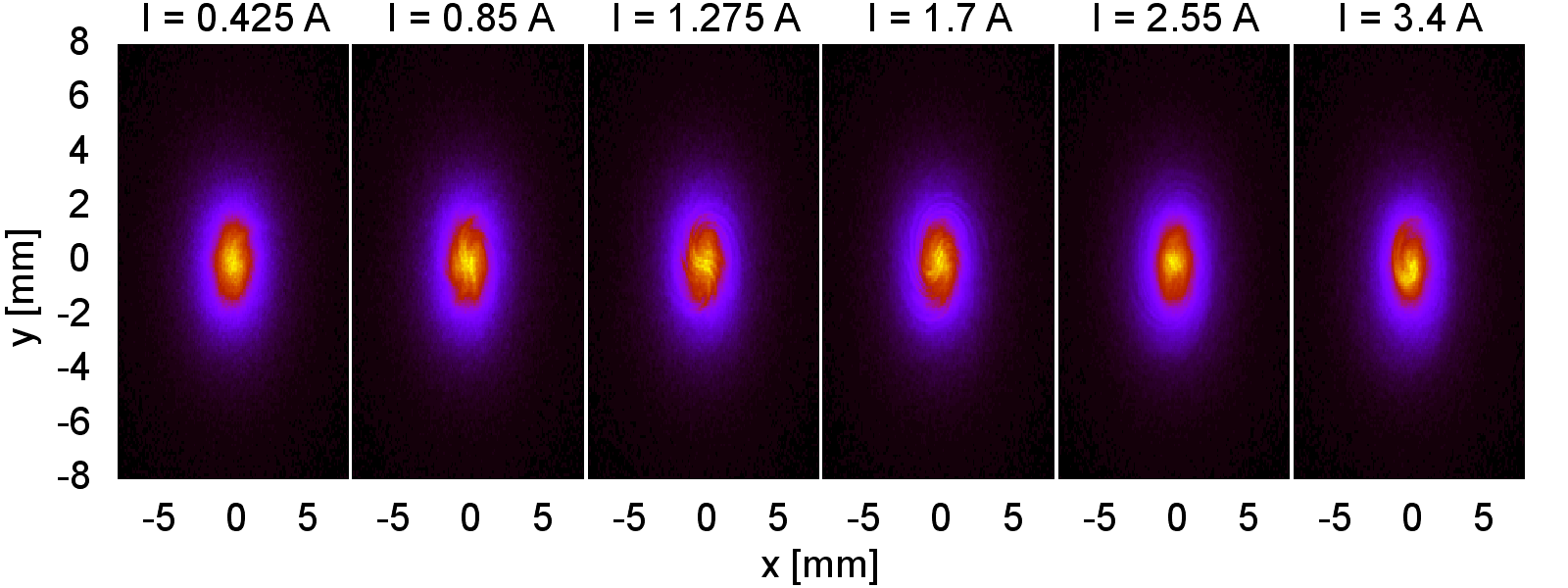}
  \caption{Current profiles at the exit of the simulation volume integrated over time. The electron gun is located towards the positive side of the plot. \label{fig:transverse_profiles_image_mm}}
\end{figure}

In Figure~\ref{fig:transverse_profiles_image_mm} the transverse profiles
of the electron beam are displayed, showing significant filamentation.
This effect is possibly of numerical origin, either from a statistical
bias in the original particle distribution or from the influence of the
solver grid on the beam, which in these simulations had no emittance.
Should fluctuations in the current density in radial direction arise at
one point, these would be converted into the observed spiral shape by
the dependence of the rotation frequency around the beam center. To
understand the origin of the fluctuations, simulations could be made for
a simpler system without bending and possibly compression.

\subsection{Calculation of kicks on the circulating beam}

As described in Ref.~\cite{stancari_kickfit}, the integrated electric field
as well as the integrated potential in the area of the electron lens
that the circulating beam in IOTA passes can be integrated and fitted
using a Chebyshev series. These kick maps can then be used as input for
long-term tracking simulations of the beam in IOTA. Using Chebyshev
polynomials has the advantage that they are orthogonal, making the
resulting coefficients independent, and that they have the same range of
values, so that the relative magnitude of the coefficients reflects
their weight. Furthermore, it is easy to express the symplecticity
condition and to require that the calculated maps are
symplectic~\cite{stancari_kickfit}, which is important for long-term
tracking.

The integrated potential as well as the kicks (integrated electric
fields) are defined as
\beq
\func{V}{x,y}=\int_{z_{0}}^{z_{1}} \mathrm{d}z\,\func{\varphi}{x,y,z} \text{ and } \func{k_{i}}{x,y}=\int_{z_{0}}^{z_{1}} \mathrm{d}z\,\func{E_{i}}{x,y,z}\mathrm{.}
\eeq
The change in angle of a passing beam particle is then given by
\(k_{x,y} / (E\rho)\) with the electric rigidity defined as \((E\rho)=p
v/e=\gamma m \beta^{2}c^{2}/e\). For a maximum kick
\(k_{x,w}=\SI{43.35}{\kilo\volt}\) (maximum value in the
\SI{1.7}{\ampere} McMillan case), the change in angle is
\SI{0.29}{\milli\radian}. The potential and the field were calculated
using bender on a grid with \SI{0.4}{\milli\meter} spacing and then
integrated over \(z_{0}=\SI{30}{\centi\meter}\) to
\(z_{1}=\SI{90}{\centi\meter}\) using the trapezoidal rule in a range of
\(\pm a_x\) and \(\pm a_y\) around the central trajectory at
\(x=\SI{42}{\micro\meter}\) and \(y=\SI{1.22}{\milli\meter}\),
\(a_x=a_y=\SI{1.2}{\centi\meter}\). The particle distributions from the
"high" resolution transport simulations were used.

The expansion of the integrated potential is defined as
\beq
\func{V}{x,y} = \sum_{n=0}^{N}\sum_{j=0}^{n} C_{j,(n-j)} \func{T_{j}}{\frac{x}{a_{x}}} \func{T_{n-j}}{\frac{y}{a_{y}}}\mathrm{.}\label{eq:intpot_expansion}
\eeq
The \(T_{j}\) are the Chebyshev polynomials of first kind and \(N\) the
chosen order of expansion. The kicks can be calculated from the
derivatives of \(\func{V}{x,y}\):
\begin{align*}
\func{k_{x}}{x,y} = -\frac{\partial V}{\partial x} = -\frac{1}{a_x}\sum_{n=0}^{N}\sum_{j=0}^{n} C_{j,(n-j)} \func{T_{j}'}{\frac{x}{a_{x}}} \func{T_{n-j}}{\frac{y}{a_{y}}}\\
\func{k_{y}}{x,y} = -\frac{\partial V}{\partial y} = -\frac{1}{a_y}\sum_{n=0}^{N}\sum_{j=0}^{n} C_{j,(n-j)} \func{T_{j}}{\frac{x}{a_{x}}} \func{T_{n-j}'}{\frac{y}{a_{y}}}
\end{align*}
For the derivatives of the Chebyshev polynomials \(\func{T_{j}'}{x}\),
the relation
\beq
\func{T_{n}'}{x} = n\func{U_{n-1}}{x} = n\begin{cases}
2\sum_{j\text{ odd}}^{n-1}\func{T_{j}}{x} & n-1\text{ odd}\\
2\sum_{j\text{ even}}^{n-1}\func{T_{j}}{x}-1 & n-1\text{ even}
\end{cases}
\eeq
was used to avoid numerical problems at \(x=0\) that occur in other
definitions.

Fits were then made using singular value decomposition of the
Vandermonde matrix, fitting either only the integrated potentials
\(\func{V}{x,y}\) or the integrated potential as well as the kicks
calculated from the electric field from the bender simulation. 

\begin{figure}
\begin{tabular}{cc}
\includegraphics[height=0.475\textwidth, angle=-90]{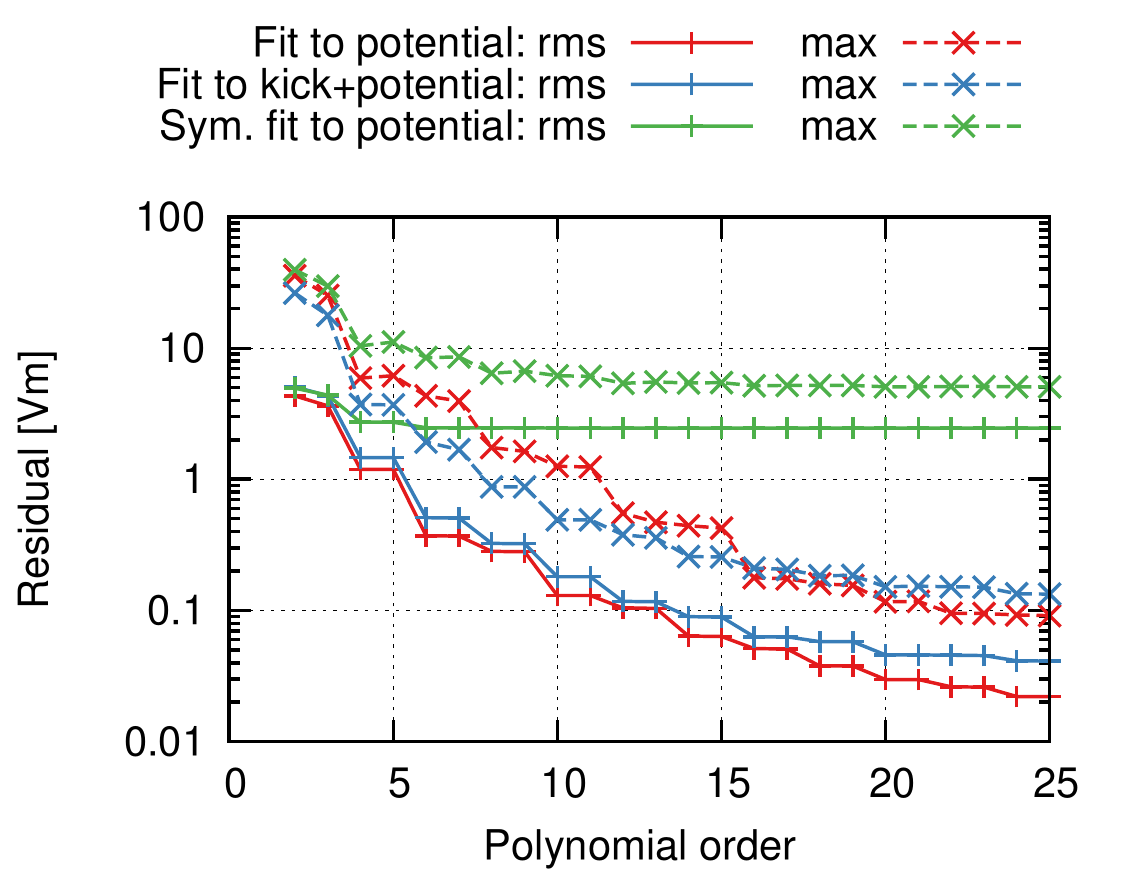} &
\includegraphics[height=0.475\textwidth, angle=-90]{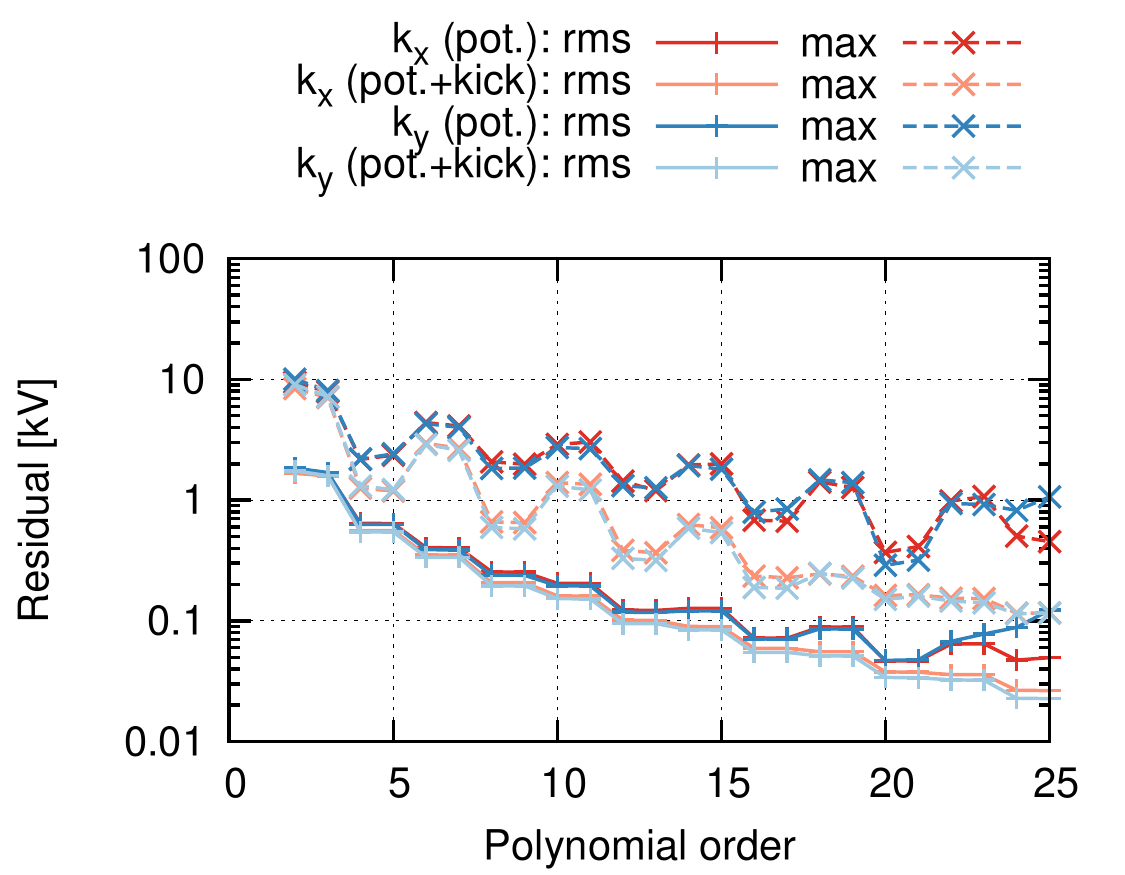} \\
\includegraphics[height=0.475\textwidth, angle=-90]{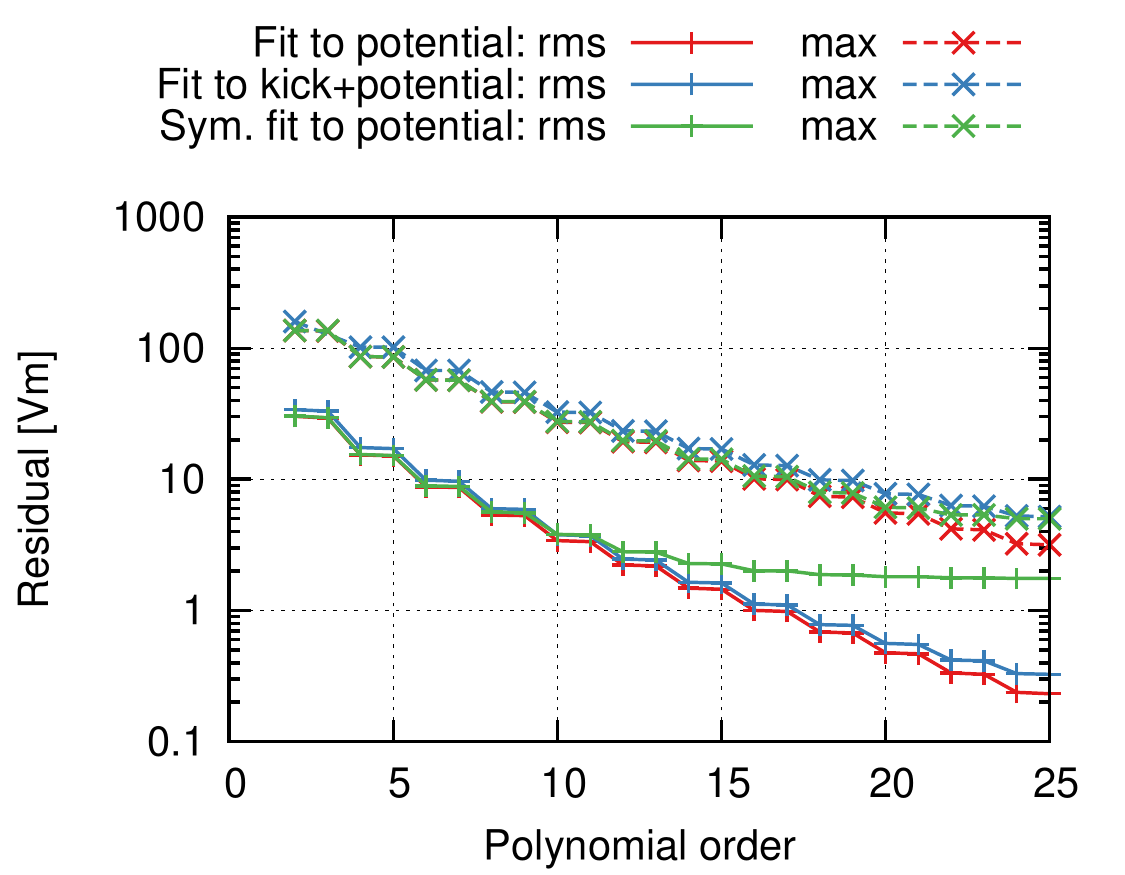} &
\includegraphics[height=0.475\textwidth, angle=-90]{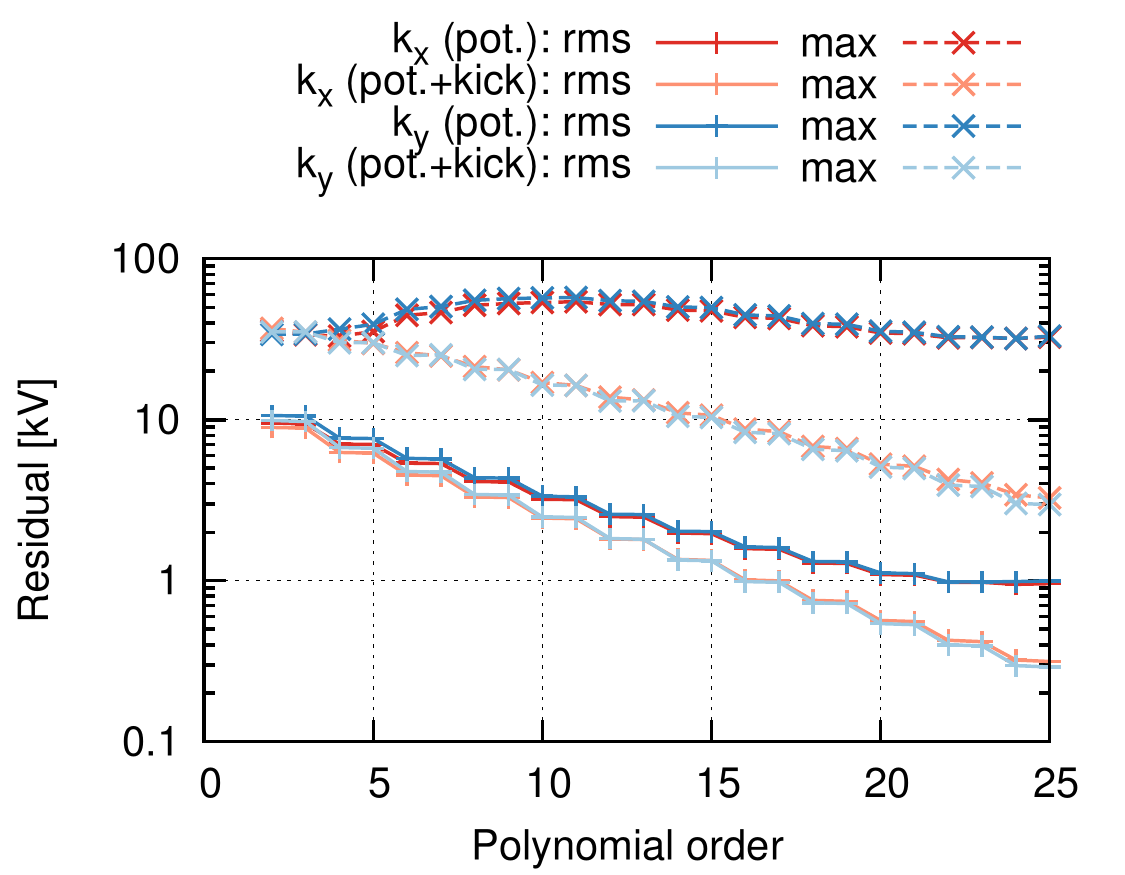} \\
Residuals of the integrated potentials & Residuals of the kicks
\end{tabular}
\caption{Errors in the fits vs. the maximum order of expansion.
Top plots: homogeneous distribution; total ranges of values:
\(V_w=\SI{341.9}{\volt\meter}\), \(k_{x,w}=\SI{19.73}{\kilo\volt}\),
\(k_{y,w}=\SI{19.0}{\kilo\volt}\).
Bottom plots: McMillan distribution;
total ranges of values: \(V_w=\SI{496.8}{\volt\meter}\),
\(k_{x,w}=\SI{43.35}{\kilo\volt}\), \(k_{y,w}=\SI{39.7}{\kilo\volt}\).}
\label{fig:error_fit}
\end{figure}

\begin{turnpage}
\begingroup
\squeezetable
\begin{table}
    \centering
    \caption{Coefficients \(C_{i,j}\) in V\(\cdot\)m, \(a_x=a_y=\SI{1.2}{\centi\meter}\) for a homogeneous beam, \(R_{\mathrm{b,gun}}=\SI{2}{\centi\meter}\), \(I=\SI{2}{\ampere}\) \label{tab:cij_kv}}
   \begin{tabular}{lrrrrrrrrrrrrrrrrrrr}
 & \(\func{T_{0}}{y/a_y}\) & \(\func{T_{1}}{y/a_y}\) & \(\func{T_{2}}{y/a_y}\) & \(\func{T_{3}}{y/a_y}\) & \(\func{T_{4}}{y/a_y}\) & \(\func{T_{5}}{y/a_y}\) & \(\func{T_{6}}{y/a_y}\) & \(\func{T_{7}}{y/a_y}\) & \(\func{T_{8}}{y/a_y}\) & \(\func{T_{9}}{y/a_y}\) & \(\func{T_{10}}{y/a_y}\) & \(\func{T_{11}}{y/a_y}\) & \(\func{T_{12}}{y/a_y}\) & \(\func{T_{13}}{y/a_y}\)\\
\(\func{T_{0}}{x/a_x}\) & -2.28e+02 & -4.21e+01 & 4.58e+01 & 1.44e+00 & -2.11e+00 & 8.84e-03 & -1.23e-01 & -6.60e-03 & -1.12e-02 & 5.87e-03 & -9.89e-03 & -2.42e-03 & 3.64e-04 & -2.22e-04\\
\(\func{T_{1}}{x/a_x}\) & -4.31e+00 & -4.02e-01 & -2.90e-01 & 2.07e-02 & 1.15e-02 & -7.69e-03 & 9.98e-03 & -4.34e-03 & -4.78e-03 & 4.82e-03 & -1.65e-04 & -1.70e-03 & 2.70e-03 & 2.40e-04\\
\(\func{T_{2}}{x/a_x}\) & 5.23e+01 & 6.19e+00 & -9.63e+00 & 1.42e-01 & -1.52e+00 & -6.54e-02 & 3.66e-01 & -2.17e-02 & 8.51e-02 & 3.85e-04 & -9.08e-04 & 1.55e-03 & 3.47e-03 & 1.06e-03\\
\(\func{T_{3}}{x/a_x}\) & -1.03e-01 & 9.66e-03 & -1.65e-02 & -1.20e-02 & 2.17e-02 & -4.09e-04 & -5.70e-04 & 1.58e-03 & -3.60e-03 & -1.34e-03 & 7.80e-04 & 3.65e-04 & -1.91e-04 & -3.28e-04\\
\(\func{T_{4}}{x/a_x}\) & -2.20e+00 & 9.26e-02 & -1.46e+00 & -9.78e-02 & 7.27e-01 & -3.81e-02 & 2.81e-01 & 2.64e-03 & -1.15e-01 & 1.10e-02 & -3.51e-02 & -2.31e-04 & 1.42e-03 & -1.28e-03\\
\(\func{T_{5}}{x/a_x}\) & -5.97e-03 & 4.76e-03 & 8.19e-03 & -8.19e-03 & 1.46e-02 & -1.62e-03 & -1.30e-02 & 5.81e-03 & 2.02e-03 & -2.07e-03 & 2.33e-03 & 3.32e-04 & -1.06e-03 & -5.20e-05\\
\(\func{T_{6}}{x/a_x}\) & -1.30e-01 & -3.80e-02 & 3.43e-01 & -2.23e-02 & 2.60e-01 & 5.90e-03 & -2.01e-01 & 1.93e-02 & -8.05e-02 & -4.92e-03 & 4.02e-02 & -2.38e-03 & 1.56e-02\\
\(\func{T_{7}}{x/a_x}\) & 1.08e-03 & -4.96e-03 & 5.84e-03 & 6.41e-03 & -7.49e-03 & 1.09e-03 & -7.44e-04 & -2.20e-03 & 3.89e-03 & 6.33e-04 & -1.44e-03 & -7.66e-05\\
\(\func{T_{8}}{x/a_x}\) & -8.86e-03 & -1.05e-03 & 6.21e-02 & 6.40e-03 & -1.00e-01 & 7.77e-03 & -7.52e-02 & -4.97e-03 & 7.11e-02 & -1.48e-03 & 5.71e-0\\
\(\func{T_{9}}{x/a_x}\) & -2.34e-03 & 1.47e-05 & 1.59e-03 & -1.10e-03 & -9.48e-04 & 2.86e-04 & 1.62e-03 & 7.25e-04 & -1.03e-03 & -6.64e-04\\
\(\func{T_{10}}{x/a_x}\) & -8.27e-03 & 1.90e-04 & 5.55e-03 & 1.24e-03 & -3.15e-02 & -1.46e-03 & 3.35e-02 & -1.14e-03 & 6.21e-03\\
\(\func{T_{11}}{x/a_x}\) & 7.64e-04 & 2.69e-03 & -1.86e-03 & -2.67e-03 & 1.39e-05 & 5.73e-04 & 1.46e-03 & -6.44e-04\\
\(\func{T_{12}}{x/a_x}\) & -4.92e-04 & 4.25e-04 & -4.45e-04 & -6.61e-04 & 9.31e-06 & -3.52e-04 & 1.54e-02\\
\(\func{T_{13}}{x/a_x}\) & 1.69e-03 & -2.05e-03 & -6.92e-04 & 2.52e-03 & -3.02e-05 & -1.22e-04 & & & & \(\func{T_{14}}{y/a_y}\) & \(\func{T_{15}}{y/a_y}\) & \(\func{T_{16}}{y/a_y}\) & \(\func{T_{17}}{y/a_y}\) & \(\func{T_{18}}{y/a_y}\)\\
\(\func{T_{14}}{x/a_x}\) & 2.38e-03 & -4.44e-05 & 1.43e-03 & -1.80e-04 & 6.50e-03 & & & & \(\func{T_{0}}{x/a_x}\) & 1.84e-03 & 1.26e-03 & 6.57e-04 & -9.98e-04 & 4.88e-04 \\
\(\func{T_{15}}{x/a_x}\) & -1.51e-03 & 2.01e-03 & 1.17e-03 & -8.09e-04 & & & & &\(\func{T_{1}}{x/a_x}\) & -2.67e-03 & 3.35e-04 & 9.77e-04 & -3.50e-04 \\
\(\func{T_{16}}{x/a_x}\) & 8.49e-05 & -2.88e-04 & -2.36e-03 & & & & & & \(\func{T_{2}}{x/a_x}\) & 2.85e-04 & -7.49e-04 & -3.76e-03\\
\(\func{T_{17}}{x/a_x}\) & 2.60e-04 & -1.10e-03 & & & & & & & \(\func{T_{3}}{x/a_x}\) & 3.10e-04 & 3.91e-04 \\
\(\func{T_{18}}{x/a_x}\) & 4.15e-04 & & & & & & & & \(\func{T_{4}}{x/a_x}\) & 5.94e-03 \\
\end{tabular}
\end{table}
\endgroup
\end{turnpage}

\begin{turnpage}
\begingroup
\squeezetable
\begin{table}
    \centering
    \caption{Coefficients \(C_{i,j}\) in V\(\cdot\)m, \(a_x=a_y=\SI{1.2}{\centi\meter}\) for a McMillan beam, \(a_{\mathrm{gun}}=\SI{3.6}{\mm}\), \(R_{\mathrm{b,gun}}=\SI{2.2}{\centi\meter}\), \(I=\SI{1.7}{\ampere}\) \label{tab:cij_mm}}
   \begin{tabular}{lrrrrrrrrrrrrrrrrrrr}
 & \(\func{T_{0}}{y/a_y}\) & \(\func{T_{1}}{y/a_y}\) & \(\func{T_{2}}{y/a_y}\) & \(\func{T_{3}}{y/a_y}\) & \(\func{T_{4}}{y/a_y}\) & \(\func{T_{5}}{y/a_y}\) & \(\func{T_{6}}{y/a_y}\) & \(\func{T_{7}}{y/a_y}\) & \(\func{T_{8}}{y/a_y}\) & \(\func{T_{9}}{y/a_y}\) & \(\func{T_{10}}{y/a_y}\) & \(\func{T_{11}}{y/a_y}\) & \(\func{T_{12}}{y/a_y}\) & \(\func{T_{13}}{y/a_y}\)\\
\(\func{T_{0}}{x/a_x}\) & -2.08e+02 & -4.22e+01 & 5.62e+01 & 3.89e+00 & -1.08e+01 & -1.01e+00 & 3.43e+00 & 4.03e-01 & -1.43e+00 & -1.81e-01 & 6.30e-01 & 1.06e-01 & -3.53e-01 & -5.11e-02\\
\(\func{T_{1}}{x/a_x}\) & -2.82e+00 & -1.54e-01 & -8.85e-01 & -1.09e-01 & 2.34e-01 & 4.38e-02 & -7.83e-02 & -1.88e-02 & 2.53e-02 & 6.64e-03 & -1.58e-02 & -3.26e-03 & 5.96e-03 & 6.22e-04\\
\(\func{T_{2}}{x/a_x}\) & 6.82e+01  & 1.41e+01  & -3.62e+01 & -3.57e+00 & 1.30e+01 & 1.42e+00 & -5.41e+00 & -6.44e-01 & 2.40e+00 & 3.48e-01 & -1.25e+00 & -1.68e-01 & 5.78e-01 & 1.13e-01\\
\(\func{T_{3}}{x/a_x}\) & -7.10e-01 & -1.08e-01 & 7.43e-01 & 1.28e-01 & -3.21e-01 & -7.19e-02 & 1.34e-01 & 3.55e-02 & -6.77e-02 & -1.93e-02 & 2.79e-02 & 8.08e-03 & -1.77e-02 & -4.18e-03\\
\(\func{T_{4}}{x/a_x}\) & -1.39e+01 & -3.63e+00 & 1.34e+01 & 1.57e+00 & -6.89e+00 & -8.09e-01 & 3.45e+00 & 4.61e-01 & -1.85e+00 & -2.41e-01 & 9.01e-01 & 1.65e-01 & -5.71e-01 & -7.89e-02\\
\(\func{T_{5}}{x/a_x}\) & 3.32e-01  & 5.92e-02  & -4.82e-01 & -9.17e-02 & 2.57e-01 & 6.01e-02 & -1.42e-01 & -3.62e-02 & 6.69e-02 & 1.87e-02 & -4.40e-02 & -1.14e-02 & 2.07e-02 & 4.70e-03\\
\(\func{T_{6}}{x/a_x}\) & 4.50e+00  & 1.29e+00  & -5.68e+00 & -6.98e-01 & 3.50e+00 & 4.57e-01 & -2.14e+00 & -2.69e-01 & 1.15e+00 & 1.96e-01 & -7.66e-01 & -9.07e-02 & 3.39e-01\\
\(\func{T_{7}}{x/a_x}\) & -1.85e-01 & -3.13e-02 & 2.89e-01 & 5.48e-02 & -1.94e-01 & -4.42e-02 & 1.06e-01 & 2.66e-02 & -7.34e-02 & -1.85e-02 & 3.18e-02 & 7.85e-03\\
\(\func{T_{8}}{x/a_x}\) & -1.87e+00 & -5.16e-01 & 2.54e+00 & 3.43e-01 & -1.89e+00 & -2.27e-01 & 1.16e+00 & 1.86e-01 & -8.47e-01 & -9.19e-02 & 3.83e-01\\
\(\func{T_{9}}{x/a_x}\) & 1.01e-01  & 1.41e-02  & -1.90e-01 & -3.46e-02 & 1.19e-01 & 2.70e-02 & -9.30e-02 & -2.18e-02 & 4.13e-02 & 9.90e-03\\
\(\func{T_{10}}{x/a_x}\) & 8.28e-01  & 2.46e-01  & -1.32e+00 & -1.52e-01 & 9.22e-01 & 1.41e-01 & -7.71e-01 & -8.10e-02 & 3.84e-01\\
\(\func{T_{11}}{x/a_x}\) & -6.79e-02 & -9.52e-03 & 1.04e-01 & 1.86e-02 & -9.33e-02 & -1.99e-02 & 4.66e-02 & 1.03e-02\\
\(\func{T_{12}}{x/a_x}\) & -4.46e-01 & -1.09e-01 & 6.09e-01 & 8.68e-02 & -5.79e-01 & -6.22e-02 & 3.42e-01\\
\(\func{T_{13}}{x/a_x}\) & 3.49e-02  & 3.89e-03  & -7.72e-02 & -1.34e-02 & 4.66e-02 & 9.27e-03 & & & & \(\func{T_{14}}{y/a_y}\) & \(\func{T_{15}}{y/a_y}\) & \(\func{T_{16}}{y/a_y}\) & \(\func{T_{17}}{y/a_y}\) & \(\func{T_{18}}{y/a_y}\)\\
\(\func{T_{14}}{x/a_x}\) & 2.09e-01  & 5.83e-02  & -3.66e-01 & -4.13e-02 & 2.70e-01 & & & & \(\func{T_{0}}{x/a_x}\) & 1.65e-01 & 3.31e-02 & -9.67e-02 & -2.18e-02 & 6.45e-02 \\
\(\func{T_{15}}{x/a_x}\) & -2.55e-02 & -2.94e-03 & 4.17e-02 & 6.96e-03 & & & & &\(\func{T_{1}}{x/a_x}\) & -2.63e-03 & -1.34e-04 & 3.21e-03 & 5.97e-04\\
\(\func{T_{16}}{x/a_x}\) & -1.19e-01 & -3.01e-02 & 1.91e-01 & & & & & & \(\func{T_{2}}{x/a_x}\) & -3.55e-01 & -6.03e-02 & 1.88e-01\\
\(\func{T_{17}}{x/a_x}\) & 1.69e-02  & 2.80e-03 & & & & & & & \(\func{T_{3}}{x/a_x}\) & 1.02e-02 & 2.21e-03 \\
\(\func{T_{18}}{x/a_x}\) & 7.26e-02 & & & & & & & & \(\func{T_{4}}{x/a_x}\) & 2.67e-01 \\
\end{tabular}
\end{table}
\endgroup
\end{turnpage}

\begin{figure}
    \centering
    \includegraphics[width=\textwidth]{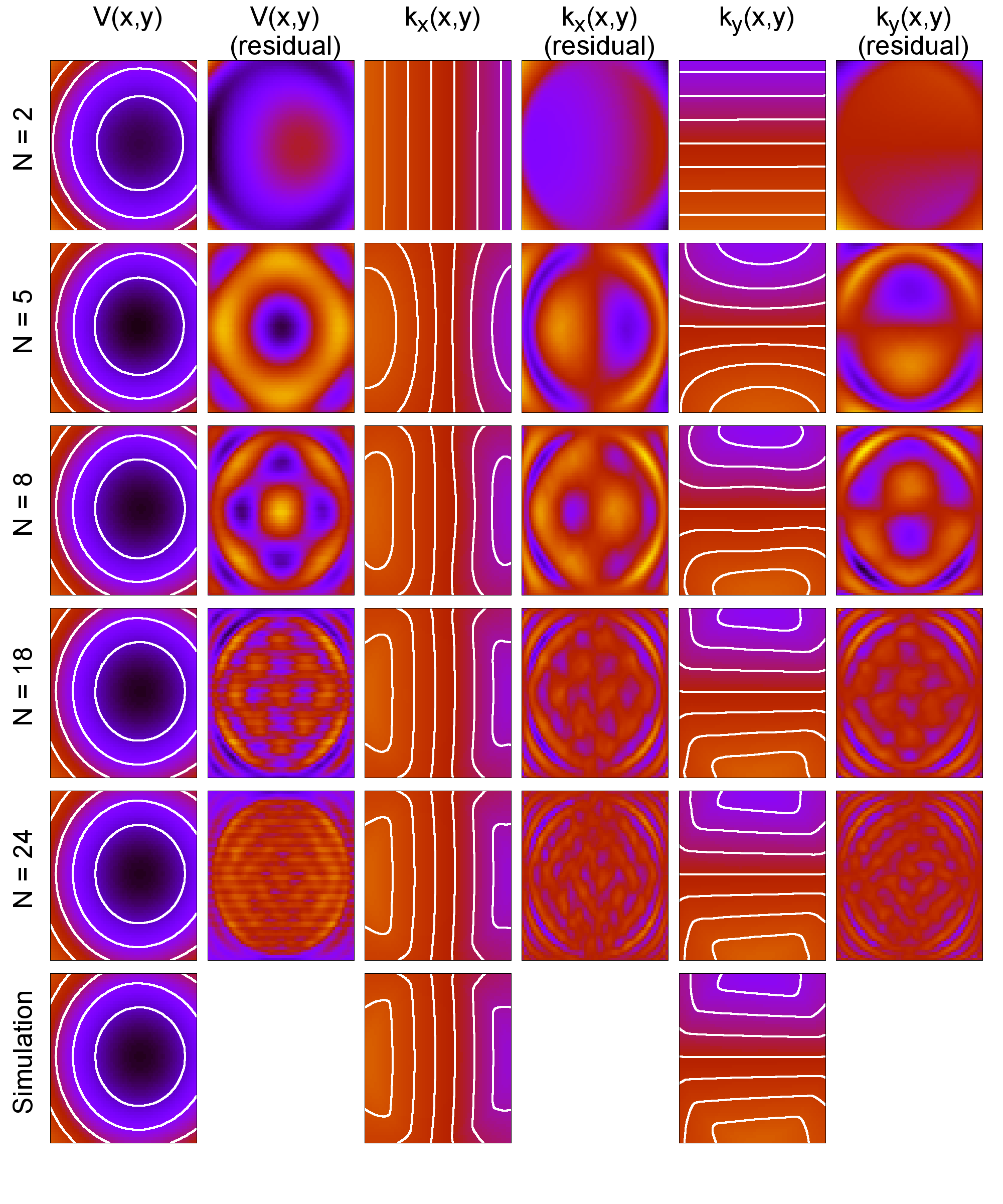}
    \caption{Integrated potentials and kicks and their residual compared to the original data from the simulation for various orders of expansion for the homogeneous beam. The parameters were calculated by fitting the integrated potential and the kicks. Contour line in the plots of the integrated potential range decrease from \SIrange{-150}{-300}{\volt\meter} from the edge of the plots to the center. For the kicks, adjacent lines have a difference of \SI{5}{\kilo\volt}. The lines in the center are at \SI{0}{\volt}.}
    \label{fig:error_fit_image_kv}
\end{figure}

\begin{figure}
    \centering
    \includegraphics[width=\textwidth]{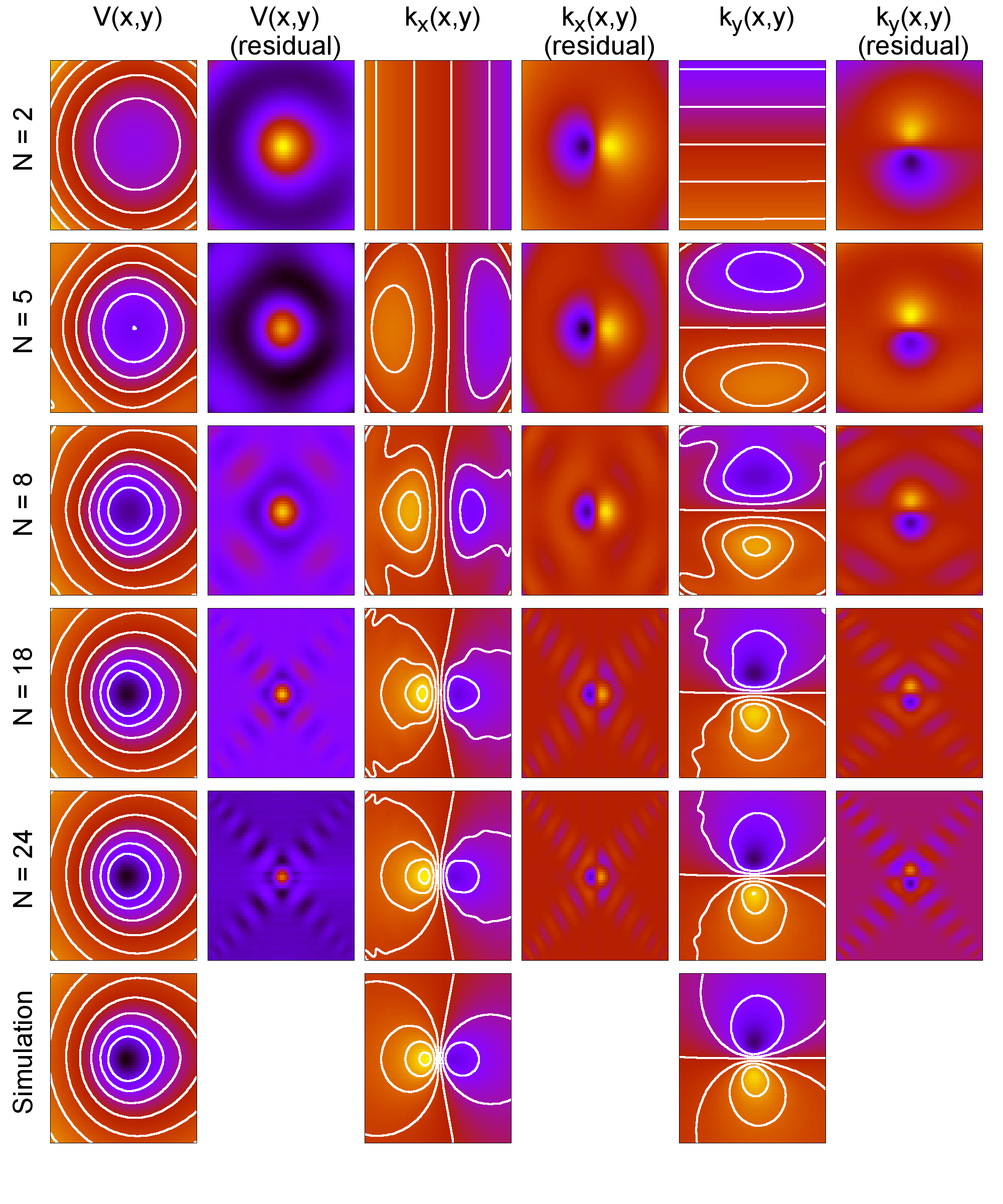}
    \caption{Integrated potentials and kicks and their residual compared to the original data from the simulation for various orders of expansion for a \SI{1.7}{\ampere}, \(a=\SI{3.6}{\milli\meter}\) McMillan distributed beam. The parameters were calculated by fitting the integrated potential and the kicks. Contour line in the plots of the integrated potential range decrease from \SIrange{-100}{-400}{\volt\meter} from the edge of the plots to the center. For the kicks, adjacent lines have a difference of \SI{10}{\kilo\volt}. The lines in the center are at \SI{0}{\volt}.}
    \label{fig:error_fit_image_mm}
\end{figure}

In Ref.~\cite{stancari_kickfit}, only the coefficients \(C_{i,j}\),
\(j\) even are fitted under the assumption that there is no vertical
distortion of the beam. As seen in section \ref{sec:sp_tracking}, the
total downward drift of a particle depends on its horizontal position.
Thus, since the beam is not completely symmetric, the fit using only
even powers in vertical direction fails to provide a good fit. This can
be seen from the green curves in Figure~\ref{fig:error_fit} (left
column). For the homogeneous distribution the effect is much more
pronounced, probably due being less peaked than the McMillan
distribution.

As was observed in Ref.~\cite{stancari_kickfit}, only fitting the integrated
potential values does not sufficiently constrain the derivatives. At
higher order, significant oscillation of the calculated kicks can be
observed. For the McMillan distribution, no convergence can be observed
in Figure~\ref{fig:error_fit} (bottom right).

For the homogenous distribution, for \(N > 18\) there is only little
change in the residuals. This is not the case for the McMillan
distribution, where the residuals continue to decrease. The reason for
this becomes apparent when looking at the distribution of the residuals
in Figure~\ref{fig:error_fit_image_mm}, which shows a clear maximum in the
residuals at the center of the beam. Due to the peaked nature of the
McMillan distribution, it takes higher contributions of higher order
polynomials to follow the peak of the distribution.

For the values of the constants \(C_{i,j}\) for the fit to both
potential and kicks to the order \(N=18\) please refer to
Tables~\ref{tab:cij_kv} and~\ref{tab:cij_mm}.

\section{Conclusion and outlook}

A design of the bending sections of the IOTA electron lens was found
using field line tracking. The sources of distortions from the magnetic
field configuration itself as well as from drifts were investigated. If
the total downward drift is canceled and the beam recentered on the
axis, the effects are in range of \SI{100}{\micro\meter}. Including
space charge forces, will result in an \(E\times B\) motion of the
particles. Its behaviour as well as limits from transverse and
longitudinal forces were estimated. Space charge calculations were then
made using \bender{}, which showed significant change in the profiles at
higher currents. Kick maps calculated from the potential and the
electric field from these simulations can now be included into tracking
simulations to investigate the influence of the asymmetric kicks
produced by the bends.

A number of improvements to the simulation as well as open questions are:
\begin{itemize}
  \item Include the electron gun or use a distribution generated from an external simulation of the electron gun to study how well the required distributions can be approximated and how the differences influence the transport through the electron lens.
  \item Parameterize various possible aberrations in the electron beam and look into their influence on the beam transport.
  \item Add the main field and the collector part of the lens to check for a good transport of the beam.
  \item Investigate the influence of geometry of the beam pipe in the bend on the kicks on the circulating beam and on electron beam transport.
  \item How large is the magnetic flux from the solenoids into the adjacent quadrupoles of IOTA?
\end{itemize}

\newpage

\appendix

\section{Electric field and potential of McMillan distributed beam}
\label{sec:mcmillan_dist}

Density (cut off at \(R_{\mathrm{b}}\)):
\beq
  \func{\rho}{r} = \frac{a^{4}}{\left(r^{2}+a^{2}\right)^{2}}\func{\theta}{r-R_{\mathrm{b}}}
\eeq

Radial integral:
\beq
\int_{0}^{r} \d r' r' \func{\rho}{r'} = -\frac{a^{4}}{2}\int_{0}^{r} \d r' \frac{\d}{\d r'}\frac{1}{r'^{2}+a^{2}} = \frac{a^{2}}{2}\left(1-\frac{a^{2}}{r^{2} + a^{2}} \right) = \frac{a^{2}}{2}\frac{r^{2}}{r^{2}+a^{2}}
\eeq

Cumulative probability density:
\begin{align*}
\func{f}{r} &= \frac{\int_{0}^{r}\d r' \func{\rho}{r'}}{\int_{0}^{R}\d r' \func{\rho}{r'}} \\
  & = \left(1+\frac{a^{2}}{c^{2}}\right)\frac{r^{2}}{r^{2}+a^{2}} \xrightarrow{c\rightarrow\infty}1-\frac{a^{2}}{r^{2}+a^{2}}=1-\sqrt{\func{\rho}{r}}
\end{align*}

Charge density (beam current \(I\), velocity \(v\)):
\beq
\func{\rho}{r} = \frac{I}{v}\frac{1}{\pi}\left(1+\frac{a^{2}}{R_{\mathrm{b}}^{2}}\right)\frac{a^{2}}{\left(r^{2}+a^{2}\right)^{2}}\func{\theta}{r-R_{\mathrm{b}}}
\eeq

Electric field (aperture \(R\)):
\beq
  \func{E_{r}}{r} = \frac{\rho_{0}}{\epsilon_{0}}\frac{1}{r} \int_{0}^{r} \d r' r' \func{\rho}{r'} = \frac{1}{2\pi\epsilon_{0}}\frac{I}{v}
  \begin{cases}
    \left(1+\frac{a^{2}}{R_{\mathrm{b}}^{2}}\right)\frac{r}{r^{2}+a^{2}} & r \leq R_{\mathrm{b}}\\
    \frac{1}{r} & R_{\mathrm{b}}\leq r < R 
  \end{cases}
\eeq

Potential:
\begin{align*}
\func{\varphi}{r}&= -\frac{1}{2\pi\epsilon_{0}}\frac{I}{v} \begin{cases}
    \func{\ln}{\frac{R_{\mathrm{b}}}{R}}+\frac{1}{2}\left(1+\frac{a^{2}}{R_{\mathrm{b}}^{2}}\right)\func{\ln}{\frac{a^{2}+r^{2}}{a^{2}+R_{\mathrm{b}}^{2}}} & r \leq R_{\mathrm{b}}\\
    \func{\ln}{\frac{r}{R}} & R_{\mathrm{b}}\leq r < R 
  \end{cases}
\end{align*}

\section{Current limit during compression of a magnetised beam due to
longitudinal space charge forces}
\label{sec:app_currentlimit}

\begin{figure}
\begin{tabular}{cc}
\includegraphics[height=0.475\textwidth, angle=-90]{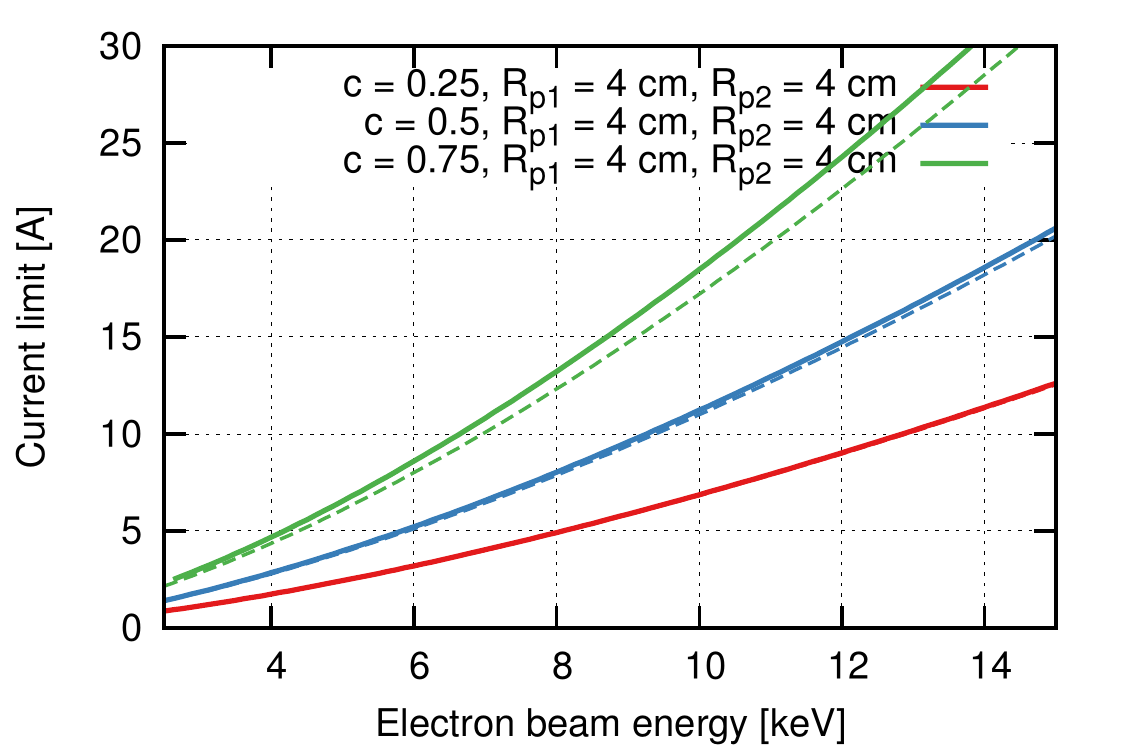} &
\includegraphics[height=0.475\textwidth, angle=-90]{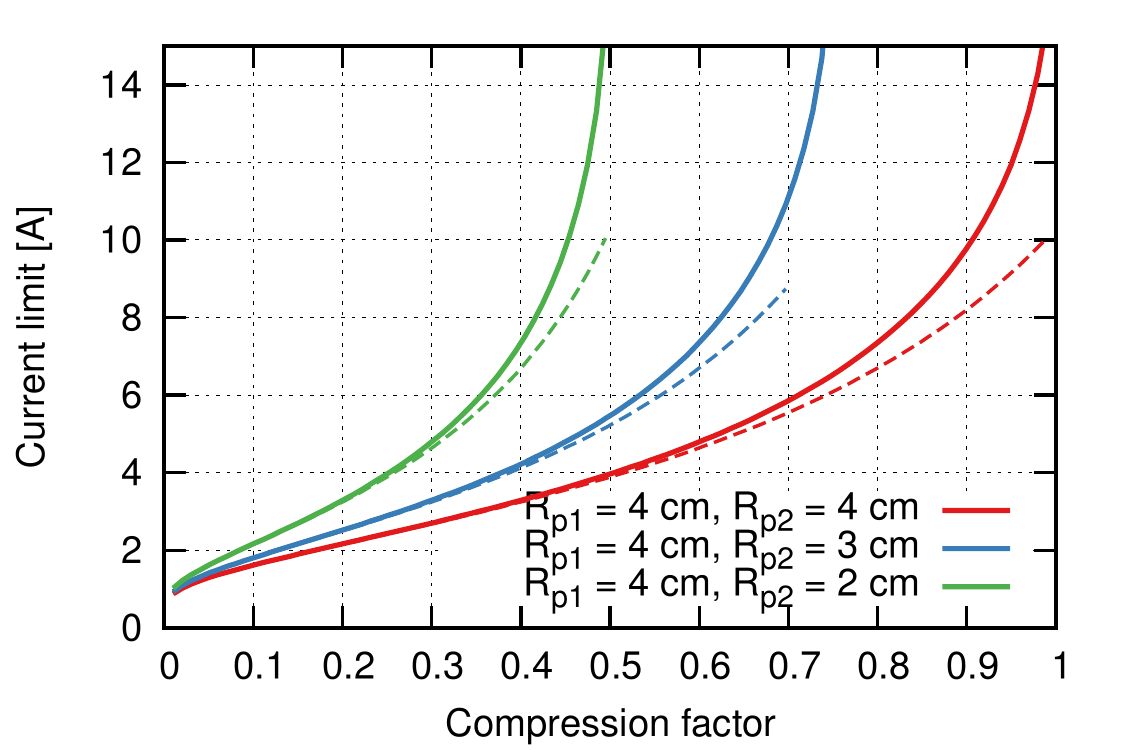}
\end{tabular}
\caption{Results of the simple current limit model. Left: Energy
dependence. Right: Current limit for different compression values at
various beam pipe radii. The solid curves are the results from the
numerical solution of the exact equation, the dashed lines are results
from the approximation.}
\label{fig:currentlimit_kv}
\end{figure}

If the Brillouin flow limit is sufficiently high, particle drifts,
additional distortions due to the magnetic field and the oscillation due
to the electric field are low, it can be assumed that the distribution
of a beam stays fixed during compression. For a homogeneous, negatively
charged beam with current \(I\), velocity \(v\) and radius
\(R_{\mathrm{b}}\) in a beam pipe of radius \(R_{\mathrm{p}}\),
\beq
\func{\varphi}{r} = \frac{1}{2\pi\epsilon_{0}}\frac{I}{v}\left[\func{\ln}{\frac{R_{\mathrm{b}}}{R_{\mathrm{p}}}}-\frac{1}{2}\left(1-\left(\frac{r}{R_{\mathrm{b}}}\right)^{2}\right)\right]\label{eq:homogeneous_potential}
\eeq
During compression by a factor of \(c < 1\) from point 1 to 2, the total
energy of each particle must be conserved, i.e.
\begin{align}
0 & = E_{2}-E_{1} = \frac{1}{2}m v_{2}^{2}-E_{1}+e\left(\func{\varphi}{r, R_{\mathrm{b}}} - \func{\varphi}{c r, c R_{\mathrm{b}}}\right)\nonumber\\
&\longrightarrow v_{2}^{3} + \left[\frac{e I}{m\pi\epsilon_{0}}\func{g}{r, R_{\mathrm{b}}, R_{\mathrm{p,1}}}\frac{1}{v_{1}} - \frac{2 E_{1}}{m}\right]v_{2} - \frac{e I}{m\pi\epsilon_{0}}\func{g}{c r, c R_{\mathrm{b}}, R_{\mathrm{p,2}}}=0\label{eq:currentlimit_long_velocity}
\end{align}
with a geometric factor
\beq
\func{g}{r, R_{\mathrm{b}}, R_{\mathrm{p}}} = \func{\ln}{\frac{R_{\mathrm{b}}}{R_{\mathrm{p,1}}}}-\frac{1}{2}\left(1-\left(\frac{r}{R_{\mathrm{b}}}\right)^{2}\right)\mathrm{,}
\eeq
which depends on the beam size \(R_{\mathrm{b}}\) as well as the radius
\(R_{\mathrm{p}}\) at which the beam pipe is placed. Equation
(\ref{eq:currentlimit_long_velocity}) assumes, that all particles move
at the same velocity by setting the velocity in the formula for the
potential to the velocity of the "test particle". Note, that when the
beam pipe follows the beam, i.e. \(R_{\mathrm{p,2}}=c
R_{\mathrm{p,1}}\), the terms depending on the current drop out if
\(v_{1}=v_{2}\). This suggests, that there is no current limit in this
case. Since a potential of the form (\ref{eq:homogeneous_potential})
ignores longitudinal beam size variation, this is probably not
completely the case. However, should a higher current be required for
future projects, using a beam pipe that follows the beam might be
helpful and should be investigated.

Equation (\ref{eq:currentlimit_long_velocity}) has a real, positive root if
\beq
  4\left(\frac{2E_{1}}{m} - \frac{e I}{\pi\epsilon_{0}}\frac{1}{\sqrt{2 m E}}\func{g}{r, R_{\mathrm{b}}, R_{\mathrm{p,1}}}\right)^{3}-27\left(\frac{e I}{m\pi\epsilon_{0}}\func{g}{ c r, c R_{\mathrm{b}}, R_{\mathrm{p,2}}}\right)^{2} > 0
\eeq
In all other cases, there is always a negative real root, which is not
of interest, as well as imaginary ones. This relation could
theoretically be solved analytically. For compactness, only the solution
to the equation which neglects the \(I^{3}\) term is given here.
\beq
I < \frac{4\pi\epsilon_{0}}{e\sqrt{m}}E^{3/2}\left(\frac{g_{1}-\sqrt{3 g_{2}^{2} - g_{1}^{2}/3}}{4g_{1}^{2}-9g_{2}^{2}}\right) = 4\pi\epsilon_{0}\sqrt{\frac{e}{m}}U^{3/2}\left(\frac{g_{1}-\sqrt{3 g_{2}^{2} - g_{1}^{2}/3}}{4g_{1}^{2}-9g_{2}^{2}}\right)\label{eq:currentlimit_long_solution2}
\eeq
A comparison between this solution and a numerical solution can be found
in Figure~\ref{fig:currentlimit_kv}. Equation
\ref{eq:currentlimit_long_solution2} provides a more conservative limit,
especially in the range of low compression, where the current limit is
probably not very relevant anyway.

The geometric factor is still dependent on the ratio \(r/
R_{\mathrm{b}}\). However, due to the assumption that all particles in
the beam have the same velocity, the values for particles other than
that on the axis are probably not reliable. The limit for the central
particles is most severe, so setting \(r=0\) should provide an
approximate lower current limit. Higher currents might still be
possible. However the equation \ref{eq:currentlimit_long_solution2}
provides an energy scaling. The energy dependence is the same as that of
Child-Langmuir law.

The estimation fails to provide a limit when \(R_{p2}/R_{p1} < c\). In
this case, the potential in the beam should be lower in the compressed
region than in the source.

\bibliography{iota-elens-sim}

\end{document}